\documentclass[useAMS,usenatbib]{mn2e}
\usepackage[english]{babel}
\usepackage[latin1]{inputenc}
\usepackage{amsmath}
\usepackage{amsfonts}
\usepackage{amssymb}
\usepackage[]{graphicx}
\usepackage{verbatim}
\usepackage{mathrsfs}
\usepackage{enumerate} 
\usepackage{paralist}
\usepackage[small, bf]{caption}

\title[Galactic and cosmic Type Ia SN rates]{Galactic and Cosmic Type Ia SN rates: is it possible to impose constraints on SNIa progenitors?}

\author[I. Bonaparte et al.]
  {I. Bonaparte,$^1$\thanks{ilariabonaparte@libero.it}
  F. Matteucci,$^{1,2,3}$ , S. Recchi,$^{4}$, E. Spitoni$^{1}$,  A. Pipino$^{5}$, V. Grieco${^1}$\\
    $^1$ Department of Physics, Trieste University, Via Valerio 2, 34100, Trieste\\
  $^2$   INAF, Via G.B. Tiepolo 11, 34100 Trieste\\
 $^{3}$ INFN, Via Valerio 2, 34100, Trieste \\
$^{4}$ Institute for Astrophysics, University of Vienna, Türkenschanzstrasse 17, 1180, Vienna, Austria \\
 $^{5}$Institut fur Astronomie, ETH Zurich, CH-8093 Zurich, Switzerland }
\date{Released 2013 Xxxxx XX}

\def\LaTeX{L\kern-.36em\raise.3ex\hbox{a}\kern-.15em
    T\kern-.1667em\lower.7ex\hbox{E}\kern-.125emX}

\begin{document}

\label{firstpage}

\maketitle

\begin{abstract}
We compute the Type Ia supernova rates in typical elliptical galaxies by varying the progenitor models for Type Ia supernovae. To do that a formalism which takes into account the delay distribution function (DTD) of the explosion times and a given star formation history is adopted. Then the chemical evolution for ellipticals with baryonic initial masses $10^{10}$, $10^{11}$ and $10^{12}M_{\odot}$ is computed, and the mass of Fe produced by each galaxy is precisely estimated. We also compute the expected Fe mass ejected by ellipticals in typical galaxy clusters (e.g. Coma and Virgo), under different assumptions about Type Ia SN progenitors. As a last step, we compute the cosmic Type Ia SN rate in an unitary volume of the Universe by adopting several cosmic star formation rates and compare it with the available and recent observational data. 
Unfortunately, no firm conclusions can be derived only from the cosmic SNIa 
rate, neither on SNIa progenitors nor on the cosmic star formation rate. Finally, by analysing all our results together, and by taking into account previous chemical evolution results, we try to constrain the best Type Ia progenitor model. We conclude that the best progenitor models for Type Ia SNe are still the single degenerate model, the double degenerate wide model, and the empirical bimodal model. All these models require the existence of prompt Type Ia supernovae, exploding in the first 100 Myr since the beginning of star formation, although their fraction should not exceed 15-20\% in order to fit chemical abundances in galaxies.
\end{abstract}


\section{Introduction}

The study of the mechanisms of supernova (SN) explosions as well as
the analysis of their nucleosynthesis products are two key ingredients
for understanding the chemical evolution of galaxies.  Type Ia
supernovae are thought to be the main contributors to the chemical
enrichment in iron in the universe and they have also a significant
influence on the early and late evolution of galaxies. In elliptical
galaxies, in fact, after the occurrence of the galactic wind and the
end of the star formation, they are the only SNe occuring. They
contribute to eject continuously energy and iron which eventually will
reach the intracluster medium (ICM). Type Ia SNe are also used to
track the Hubble law and therefore they are fundamentally important in
cosmology.
 
Therefore, the description of the evolution of the SN Ia rate, in
galaxies and in a unitary volume of the universe (cosmic rate), is a
crucial information for galaxy evolution and cosmology. The
computation of the Type Ia SN rate is related to the nature of the
progenitor systems, which unfortunately are still poorly known.  The
observed features of SNe Ia suggest that these objects may originate
from the thermonuclear explosion of CO white dwarfs (WD) of
Chandrasekhar mass and the two theoretical main scenarios which have
been proposed so are: a) the Single Degenerate (SD) scenario and b)
the Double Degenerate (DD) scenario.  Recently, it has been suggested
also a double detonation in sub-Chandrasekhar masses as a possible
mechanism for explaining some of the SNe Ia. In fact, in the last
years it has became more and more evident the existence of a variety
of Type Ia SNe. Here we will analyze only the classical DD and SD
scenarios (see Hillebrandt et al. 2013 for a recent review on all
possible scenarios).

In the original SD scenario there is a binary system in which the
primary component has mass in the range $(2-8)M_\odot$ while the
secondary component is a non-degenerate companion, a red giant or a
main sequence star (e.g. Whelan \& Iben 1973), that has mass in the
range $(0.8-8)M_\odot$. The lower limit, in the range of masses, is
due to the fact that the only systems of interest are those capable of
generating a Type Ia SN in a Hubble time to explain the existence of
SNe Ia in ellipticals. The upper limit instead is imposed by the fact
that  single stars with masses $M>8M_\odot$ ignite carbon in a
non-degenerate core and do not end their lives as CO WDs. When the
secondary star evolves and fills its Roche lobe, the WD accretes
material. Thanks to the accretion of matter, via mass transfer from
the non-degenerate companion, the primary star reaches the
Chandrasekhar mass and explodes. For many years the only
  suggested explosion mechanism was C-deflagration, but recently it
  has been suggested that in SD scenario SNe explode via a detonation,
  after deflagration has been initiated. On the other hand, for the DD
  scenario the explosion occurs via a prompt detonation or a double
  detonation (e.g. Pakmor et al. 2013). One of the limitations of the
SD scenario is that the accretion rate should be defined in a quite
narrow range of values.  To avoid this problem, Kobayashi \&
al. (1998) had proposed a similar scenario, based on the model of
Hachisu \& al. (1999), where the companion can be either a red giant
or a main sequence star, but including a metallicity effect which
suggests that no Type Ia systems can form for [Fe/H]$< -1.0$ dex. This
is due to the development of a strong radiative wind from the C-O WD
which stabilizes the accretion from the companion,  and allowing
  the WD to reach the Chandrasekhar limit for a wide binary parameter
  space than the previous scenario. This scenario will not be
considered here since models of galactic chemical evolution
(e.g. Matteucci \& Recchi, 2001) have demonstrated that the SN Ia rate
in the Galaxy reaches a maximum when [Fe/H]= -1.0 dex, thus making the
long delay due to the metallicity effects unrealistic.

In the original DD scenario the binary system is composed by two CO
WDs that, because of the emission of gravitational wave radiation,
lose angular momentum and merge achieving the Chandrasekhar mass
(e.g. Iben \& Tutukov 1984) and explode as mentioned above. The
progenitor masses are defined in the range $(5-8)M_\odot$ to ensure
two WDs of $0.7M_\odot$ and then reach the Chandrasekhar mass. The
time of the explosion is the lifetime of the secondary star plus the
time necessary to merge. The validity of this scenario requires that
the two CO WDs have an initial separation less than $3R_\odot$,
condition that can be reached by means of two different precursor
systems: a close binary and a wide binary system. The two scenarios
differ for their efficiency of the common envelope phase during the
first mass transfer, and therefore for the separation attained at the
end of the first common envelope phase.  The analysis of the SN Ia
rates in the past years has been the subject of several works, as the
pioneering works of Greggio \& Renzini (1983), Iben \& Tutukov (1984),
Tornamb\'e \& Matteucci (1986), Matteucci \& Greggio (1986), and of
the most recent works like those of Matteucci \& Recchi (2001),
Strolger \& al. (2004), Greggio (2005), Mannucci \& al. (2006),
Pritchet \& al. (2008), Totani \& al. (2008), Valiante \&
al. (2009). An extensive review on the subject can be found in Maoz \&
Mannucci (2012).  We aim at testing various distributions of explosion
times and also different star formation histories for elliptical
galaxies. In fact, once established the nature of the SN Ia
progenitors, the Type Ia SN rate is the convolution of the
distribution of the explosion times, usually called the delay time
distribution function (DTD), with the star formation rate (SFR) of the
studied galaxy.  The adopted DTDs, which describe the rate at which
SNe Ia explode as a function of time in a simple stellar population (a
starburt), refer to the single degenerate model (white dwarf plus red
giant companion) and to the double degenerate model (two white dwarfs)
as well as to empirical DTDs, as suggested by various authors. A large
convergence is found for an empirical DTD proportional to $t^{-1}$
which provides a behavior very similar to that predicted by the double
degenerate scenario (see later).  The purpose of this work is also the
computation of the cosmic Type Ia supernova rates, that is the rate as
a function of redshift of Type Ia SN explosions in a unitary volume of
the universe. The cosmic Type Ia supernova rate is computed using the
same DTDs and different cosmic star formation histories. In this case
we need to use a cosmic star formation rate that is usually expressed
in $M_{\odot}yr^{-1}Mpc^{-3}$. Such histories are partly derived from
a fit of observational data and partly from theoretical models making
different assumptions about the number density evolution of
galaxies. In fact, in the pure luminosity evolution scenario, the
number density of galaxies is considered constant as a function of
redshift. In the hierarchical galaxy formation scenario, instead, the
number density of galaxies is assumed to change with the cosmic
time. Therefore, one can in principle, put constraints on the
mechanisms of galaxy formation by comparing theoretical results with
data of cosmic star formation rate.  In this paper we will compare our
predicted cosmic Type Ia SN rates with the latest compilation of data
for the SNe Ia.  Previous works (Maoz \& Gal-Yam, 2004; Forster \&
al. 2006; Valiante \& al. 2009; Maoz \& al. 2010) have already tried
to infer constraints on SNIa progenitors from the cosmic SNIa rate. In
particular, Maoz \& al. (2010) studied Type Ia SN rates in clusters of
galaxies and concluded that a DTD $\propto$ $t^{-1/2}$ can best
reproduce the data. However, no firm conclusions were suggested.  One
difference with the previous works is that here we test the various
DTDs also in the chemical enrichment of the intracluster medium (ICM).
The paper is organized as follows: in section 2 we describe how to
compute the Type Ia SN rate in galaxies, as well as the different DTD
functions for the explosion times. In Sections 3 and 4 we present the
results for the Type Ia SN rates in a typical elliptical under
different assumptions about the SNIa progenitor models. In Section 5
we show the effects of different DTDs in galaxy models on the
predicted Fe and gas mass produced by ellipticals in galaxy clusters.
In Section 6 we describe how to compute the cosmic Type Ia SN rate by
means of different cosmic star formation rates (CSFRs) and we compare
our model results to data. In Section 7 some conclusions are drawn.

\section{Type Ia SN rate: the formalism}

The Type Ia SN rate, from a theoretical point of view, is difficult to
derive because of the uncertainty in the nature of progenitors.
Assuming that the SD scenario is valid, the SN Ia rate can be written
as (Greggio \& Renzini, 1983):

\begin{equation}
R_{SNIa}={A_{Ia}}\int_{M_{Bm}}^{M_{BM}}{\varphi(m)[\int_{\mu_Bmin}^{0.5} {f(\mu_B)\psi(t-\tau_{m2})d\mu_B]}dm}
\end{equation}
where:
\begin{itemize}
\item $\psi(t-\tau_{M2})$ is the SFR at the time of the birth of the binary system.
 \item $A_{Ia}$ is the fraction of binary systems which can give rise to Type Ia SNe only relative to the mass range $(3-16)M_\odot$;
 \item $M_{Bm}$ and $M_{BM}$ are the maximum and minimum total mass of
   the binary systems able to reproduce a SN Ia explosion. The value
   of the upper limit $M_{BM}=16M_\odot$ is due to the assumption that
   the more massive system should be made of two stars of $8M_\odot$
   each. Instead the minimum total mass of the binary system is
   assumed to be $M_{Bm}=3M_\odot$ to ensure that the companion of the
   WD is large enough to allow the WD with the minimum possible mass
   ($\sim0.5M_\odot$) to reach the Chandrasekhar mass limit ($\sim
   1.44 M_{\odot}$) after accretion.
 \item The function $f(\mu_B)$ is the distribution of the mass
   fraction of the secondary stars in binary systems, namely,
   ${\mu_B}=\frac{M2}{(M1+M2)}$ , with M1 and M2 being the primary and
   secondary masses of the system, respectively. This function is
   derived observationally and in the literature it has often been
   written  (see Greggio \& Renzini, 1983: Matteucci \& Greggio,
     1986) using the following expression:
 
\begin{equation}
f(\mu_B)=2^{1+\gamma}(1+\gamma){\mu_B}^\gamma,
\end{equation}
for $0>\mu_B\geq\frac{1}{2}$.
\item $\tau_{M2}$ is the lifetime of the secondary star and represents the time elapsed between the formation of the binary system and its explosion.
\end{itemize}
If we analyze the DD model, the supernova rate can be computed in the following way (Tornamb\'e \& Matteucci 1986);

\begin{equation}
R_{SNIa}=Cq\int_{M_{min}}^{M_{max}}\varphi(M)[\int_{S_{B_{min}}}^{S_{B_{max}}}\Psi(t-\tau_{M2} -\tau_{gw}) dlogS_B]dM
\end{equation}
where:
\begin{itemize}
 \item C is a normalization constant;
 \item $q=\frac{M_2}{M_1}=1$ is the ratio between the secondary and
   the primary mass which, in this scenario, is assumed, for
   simplicity, to be equal to unity;
 \item $S_B$ is the initial separation of the binary system at the beginning of the gravitational wave emission;
 \item $\tau_{grav}$ is the gravitational time-delay, that for systems which can give rise to SNe of Type Ia, varies from $10^6$ to $10^{10}$ years and more (Landau \& Lifshitz 1962):
\end{itemize} 

\begin{equation}
\tau_{grav}=1.48\cdot10^8{\frac{({S_B \over r_{\odot}})^4}{({M_1 \over M_{\odot}}) \cdot 
({M_2 \over M_{\odot}})\cdot ({M_1 \over M_{\odot}}+ {M_2 \over M_{\odot}})}}yr
\end{equation}

It is now possible to use a new more general formulation for the SN Ia rate, as proposed by Greggio (2005, G05), in which the Type Ia SN rate is defined like the convolution between:
\begin{compactenum}[i)]
 \item the distribution of the explosion times of SN Ia progenitors,
   that is the time delay distribution function (DTD) and
   characterizes the progenitor model;
 \item the star formation rate, namely the amount of gas turning into stars per unit time, that usually is expressed in $\frac{M_\odot}{yr}$.
\end{compactenum}
The peculiarity of this formulation is that it allows us to include any DTD that can represents a different scenarios with respect to the SD and DD:

\begin{equation}
R_{SNIa}(t)={k_\alpha}\int_{\tau_i}^{min(t,\tau_x)}{A_{Ia}}(t-\tau)\psi(t-\tau)DTD(\tau)d\tau
\end{equation}
where :
\begin{itemize}
 \item $A_{Ia}(t-\tau)$ is the fraction of binary systems which can
   give rise to Type Ia SNe and in principle it can vary in time. In
   this new formulation $A_{Ia}$ is relative to the whole range of
   star masses $(0.1-100) M_\odot$, not only relative to the mass
   range $(3-16) M_\odot$, as it is in the old formulation. This is
   unfortunately a free parameter which is fixed by reproducing the
   present time SN Ia rate.
 \item The DTD is defined in the range $({\tau_i}, {\tau_x})$ and normalized as:

\begin{equation}
\int_{\tau_i}^{\tau_x}DTD(\tau)d\tau=1
\end{equation}

 \item $\tau_i$ is the time at which the first SNe Ia start exploding, namely is the minimum delay time for the occurrence of Type Ia SNe, while $\tau_x$ is the maximum delay time;
 \item $k_\alpha$ is the number of stars per unit mass in a stellar generation and contains the IMF. In particular: 

\begin{equation}
k_\alpha=\int_{M_L}^{M_U}\varphi(m)dm
\end{equation}
with $M_L=0.1M_\odot$ and $M_U=100M_\odot$.
\end{itemize}
The IMF is the mass distribution of stars at birth and  its most common parameterization is that of Salpeter (1955) which is a simple one-slope power law generally defined in the mass range $(0.1-100)M_\odot$

\begin{equation}
\varphi(m)=am^{-(1+x)}
\end{equation}
where:
\begin{itemize}
 \item a is the normalization constant derived by imposing that 

\begin{equation}
\int_{0.1}^{100}m\varphi(m)dm=1.
\end{equation}

\end{itemize}

In this paper we have analyzed different DTDs, that are: i)the DTD of
the SD scenario as proposed by Matteucci \& Recchi (2001); ii) the DTD
of the wide DD scenario as proposed by Greggio (2005), and four
empirical DTDs which are: i) a bimodal DTD proposed by Mannucci \&
al. (2006); ii) a gaussian DTD proposed by Strolger \& al. (2004);
iii) two power law DTDs proposed by Pritchet \& al. (2008) and Totani
\& al. (2008).  It should be noted that in the literature there
  are many DTDs calculated by means of detailed binary evolution
  calculations (e.g. Yungelson \& Livio, 2000; Han \& Podsiadlowski,
  2004;Belczynski et al. 2005; Ruiter \& al. 2009; Mennekens \&
  al. 2010; Wang \& al. 2010). However, these detailed calculations
  are not always easy to use for the kind of semi-analytical
  calculations performed in this paper.

\subsection{The DTD of the SD scenario}

The SD model (Whelan \& Iben 1973), as described by Matteucci \& Recchi (MR01), is computed adopting the following formalism:

\begin{equation}
DTD(\tau)\propto\widetilde{\phi}(M_2)\frac{dM_2}{dt}
\end{equation}
which corresponds to the SN Ia rate for an instantaneous starburts. The function $\widetilde{\phi}(M_2)$ is the mass function of the secondary component that, for this DTD, is equal to:

\begin{equation}
\widetilde{\phi}(M_2)=2^{(1+\gamma)}(1+\gamma){M_2^\gamma}\frac{({M_b^{(-s-\gamma)}}-{M_B^{(-s-\gamma)}})}{(-s-\gamma)},
\end{equation}
where $s=1+x$ with $x$ being the Salpeter index. 
The derivative $\frac{dM_2}{dt}$ was obtained adopting the inverse of the formula (Padovani \& Matteucci, 1993):

\begin{equation}
\tau(M)=10^{\frac{1.338-\sqrt{1.79-0.2232(7.764-log(M))}}{0.1116}},
\end{equation}
which defines the relation between the stellar mass M expressed in
$M_\odot$ and the Main Sequence lifetime $\tau$ expressed in yr.  The
values of the parameters we will adopt in this study are $\gamma=0.5$,
$s=2.35$, $M_b=max(2M_2,M_{min})$, $M_B=M_2+0.5(M_{max})$,
$M_{min}=3M_\odot$, $M_{max}=16M_\odot$. The fraction of prompt SNe in
this scenario is $\sim 10\%$.   It is worth noting that this MR01
  DTD behaves like $t^{-1.6}$ at late times, in agreement with the
  empirical DTD found for galaxy clusters derived by Sand \&
  al. (2012).

\subsection{The DTD of the DD scenario}

The analytic formulation to describe the DTD in the DD scenario was
proposed by Greggio (2005). In that paper there is the analytic
formulation of the DTD for two different schemes, : i) close DD
scheme, in which the close binary evolution leads to a narrow
distribution of the separations, so that the initially closest
binaries merge in a short time, and the initially widest binaries tend
to populate the long $\tau_{grav}$ tail of the distribution. In
addition, the most massive binaries tend to end up with the smallest
final separation, hence merge more quickly. The ii) wide DD scheme, in
which the close binary evolution produces a wide distribution of
separations and total binary masses, and these two variables are
virtually independent. We considered only the wide DD scheme because
it is more in agreement with the observations suggesting the existence
of Type Ia SNe exploding at the present in elliptical galaxies, which
stopped to form stars several Gyrs ago.

\begin{eqnarray}
DTD(\tau)= 
\int_{\tau_{n,inf}}^{min({\tau_{n,x}},\tau)}n(\tau_n)S^C(\tau, \tau_n)d\tau_n \\ 
for\,\, CLOSE\,\, DD\\
\int_{\tau_{n,i}}^{min({\tau_{n,x}},\tau)}n(\tau_n)S^W(\tau, \tau_n)d\tau_n \\ 
for\,\, WIDE\,\, DD
\end{eqnarray}

where:

\begin{eqnarray}
S^C(\tau, \tau_n)=
\frac{(\tau-\tau_n)^\beta_g}{{\tau_{gw,x}^{(1+\beta_g)}}-{\tau_{gw,i}^{(1+\beta_g)}}}\\
for \,\, \tau_n \leq\tau-\tau_{gw,i} \\
=0 \,\, for \,\, \tau_n \geq\tau-\tau_{gw,i}\\
\end{eqnarray}

\begin{eqnarray}
S^W(\tau, \tau_n)=
f_{1,2}^W(\tau, \tau_n)^{(-0.75+0.25\beta_a)}\\ 
for \,\, \tau_n \leq\tau-\tau_{gw,i}  \\
=0 \,\,for \,\, \tau_n \geq \tau-\tau_{gw,i}
\end{eqnarray} 

The functions $S^C(\tau, \tau_n)$ and $S^W(\tau, \tau_n)$ are $\propto\frac{\partial g(\tau, \tau_n)}{\partial \tau}$ where $g(\tau, \tau_n)$ is the fraction of systems which, having a nuclear delay equal to $\tau_n$ have also a total delay shorter than $\tau$.

\begin{eqnarray}
\tau_{n,inf}=
\tau_{n,i} & \mbox  {for} & \tau<\tau_{n,i}-{\tau_{gw,x}}(\tau_{n,i})\\
\tau_n^\star & \mbox  {for} & \tau\geq\tau_{n,i}-{\tau_{gw,x}}(\tau_{n,i})
\end{eqnarray}

and 
\begin{description}
 \item$\tau_n^\star$ is the solution of the equation $\tau=\tau_n+{\tau_{gw,x}}(\tau_n)$;
 \item$\tau_n$ is the nuclear lifetime of the secondary;
 \item$\tau_i$ is the absolute minimum delay of SN Ia progenitors;
 \item$\tau_{gw}=\frac{0.15 A^4}{({m_{1R}}+{m_{2R}}){m_{1R}}{m_{2R}}}Gyr$ 
with $A, m_{1R}, m_{2R}$ respectively the separation and component masses of DD system, in solar units.
\end{description}

\subsection{Empirical DTDs}

\subsubsection{Bimodal DTD}

The bimodal DTD proposed by Mannucci \& al. (2006) (MVP06) is derived empirically thanks to the observations of supernovae in radio galaxies. The bimodal trend is given by the sum of two distinct functions:
\begin{compactenum}[i)]
 \item a prompt Gaussian centered at $5\times10^{7}yr$;
 \item a much slower function, either another Gaussian or an exponentially declining function.
\end{compactenum}
This DTD describes a situation where a percentage form $35\%$ to
$50\%$ of all SNe Ia explode during the first 100 Myr since the
beginning of star formation (prompt SNe Ia), while the rest explodes
with larger delays as long as the Hubble time and more (tardy SNe
Ia). This DTD can be analytically approximated by the following
expressions (Matteucci \& al. 2006):
  
\begin{eqnarray}
log DTD(\tau)= 1.4-50(log\tau-7.7)^2   &  \mbox{for} & \tau<85 Myr \\
-0.8-0.9(log\tau-8.7)^2 &  \mbox{for} & \tau>85 Myr 
\end{eqnarray}

This bimodal DTD can be associated to the SD scenario.

\subsubsection{Gaussian DTD}

The gaussian DTD proposed by Strolger \& al. (2004, S04) is derived
thanks to the analysis of the data obtained with Hubble Space
Telescope. These observations have allowed to discover 42 SNe in the
redshift range $0.2<z<1.6$. As these data span a large range in
redshift, they are ideal for testing the validity of Type Ia supernova
progenitor models with the distribution of expected delay times. The
result suggests that the models that requires a large fraction of
prompt SNe Ia poorly reproduces the observed redshift distribution and
are rejected at greater than $95\%$ confidence. The conclusion is that
Gaussian models best fit the observed data for mean delay times in the
range of $(2-4) Gyr$.  The formula of this DTD is:

\begin{equation}
DTD(\tau ,t_d)=\frac{1}{\sqrt{2\pi\sigma_{\tau}^2}}e^{\frac{-(\tau -t_d)^2}{2\sigma_{t_d}^2}}
\end{equation}
where $\sigma_{t_d}=0.5 t_d$ and $t_d$ is the characteristic delay time, here assumed to be 4 Gyr.

\subsubsection{Power law DTDs}

The two power law DTDs are proposed by Totani \& al. (2008) and Pritchet \& al. (2008). The relation relative to these functions is:

\begin{equation}
DTD(\tau)\propto{\tau^{\alpha}}
\end{equation}
with $\alpha=- 1$ in the Totani \& al. case, and $\alpha=- 0.5$ 
in the Pritchet \& al. case.

The DTD of Totani et al. (2008) was suggested on the basis of faint
variable supernovae detected in the Subaru/XMM-Newton Deep Survey
(SXDS). The sample used for the definition of this DTD was composed by
65 SNe showing significant spatial offset from the nuclei of the host
galaxies having an old stellar population at z $\sim 0.4-1.2$ out of
more than 1000 SXDS variable objects. This DTD supports the idea of
the DD model (Maoz \& Mannucci, 2012).  The DTD proposed by Pritchet
et al. (2008) instead was obtained by the data derived from the SNLS
survey, that is the SuperNova Legacy Survey, (Sullivan \& al. 2006).

\subsection{Properties of the different DTD functions}

\begin{figure}
\begin{center}
\includegraphics[scale=.40]{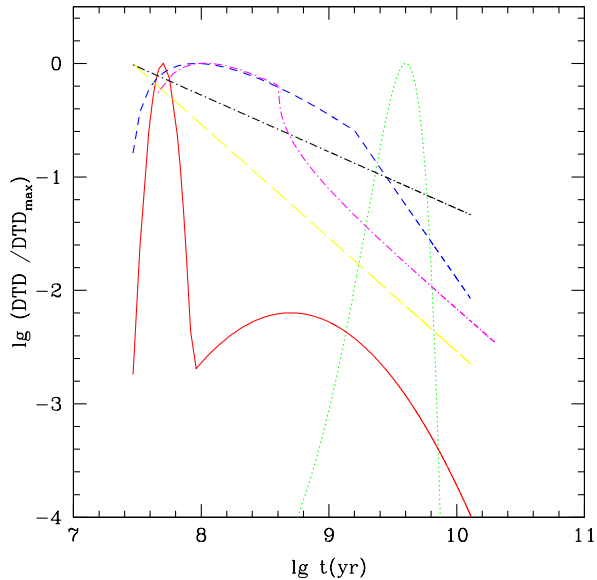}
\caption{Illustration of the various DTD functions normalized to their
  own maximum value. The solid red line is the DTD of MVP06; the
  dashed blue line is the DTD of MR01; the short dashed-dotted magenta
  line is the DTD of G05; the dotted green line is the DTD of S04; the
  black short dashed-dotted black line is the DTD of Pritchet \&
  al. (2008); the long dashed yellow line is the DTD of Totani \&
  al. (2008).}
\label{DTDs} 
\end{center}
\end{figure}

All the trends of the six different DTDs analyzed here are reported in Figure \ref{DTDs}.

The SD and DD formulations follow from general considerations on the evolutionary behavior of stars in binary systems. Some parameters play a key role in shaping the SD and DD DTDs, most notably:
\begin{enumerate}
 \item the mass range of the secondaries in systems which provide SNe Ia events;
 \item the minimum mass of the primary which yields a massive enough CO WD to ensure the explosion; 
\item the efficiency of the common envelope process;
 \item the efficiency of accretion, for the SD model; 
 \item the distribution of the separation, for the DD system at their birth.
\end{enumerate}
The empirical DTDs instead are derived directly from observations.

\section{The predicted Type Ia rates in a typical elliptical}

We have predicted the Type Ia SN rate for an elliptical galaxy having
a baryonic mass of $10^{11}M_{\odot}$. The assumed cosmology, through
all the paper, is the Lambda Cold Dark Matter ($\Lambda$CDM) cosmology
with $\Omega_M=0.3$, $\Omega_{\Lambda}=0.7$ and $H_0\sim 67 Km
s^{-1}Mpc^{-3}$.  The SN rate can be computed like the convolution of
a given SFR, that it defined as the amount of gas turning into stars
per unit time, and a particular DTD, that is the distribution of the
explosion times, as shown in eq. (5).

\begin{figure}
\begin{center}
\includegraphics[scale=.40]{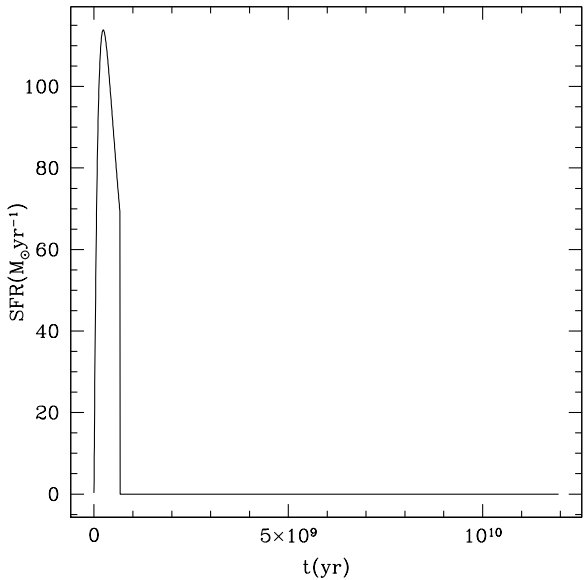}
\caption{The SFR for an elliptical galaxy of $10^{11}M_{\odot}$. The occurrence of a galactic wind at $t_{GW}=0.67$ Gyr stops the SF.}
\label{SFR} 
\end{center}
\end{figure}

Initially, with the purpose of verifying mostly the dependence of the
SN rates on the DTD, we have computed the Type Ia SN rate using only a
given SFR and the six different distribution functions studied
previously.  The trend of this SFR is shown in Figure \ref{SFR}. This
SFR is obtained using the Pipino \& Matteucci (2004) model applied to
an elliptical galaxy of $10^{11} M_{\odot}$ of initial luminous mass,
when the SD model for SN Ia progenitors is assumed. In Pipino \&
Matteucci (2004) model the SFR stops when a galactic wind, triggered
mainly by SNe II and Ib,c and partly by SNe Ia, occurs.  Other
  mechanisms exist to devoid galaxies from gas such as ram pressure
  stripping, tidal stirring and strangulation, but we do not consider
  them here. In any case, there should be a mechanism that quenches
  star formation in ellipticals and galactic winds seem the most
  reasonable one. Moreover, as shown by Pipino \& Matteucci (2004),
  galactic winds occurring first in massive than in small ellipticals
  can very well explain the increase of the [$\alpha$/Fe] ratio with
  galactic mass in ellipticals. Whether such a wind continues for the
  whole galaxy lifetime is not clear and probably the other mechanisms
  will be at action as well. However, what matters here is that the
  gas is soon or later lost into the ICM.

As it can be seen from the Figure \ref{SFR} the SFR has a simple form
given by the Schmidt-Kennicut law, $\psi(t)=\nu\sigma_{gas}^k$, with
an efficiency of star formation $\nu=10 Gyr^{-1}$ and $k=1$. The SFR
is halted by the occurrence of a galactic wind at $t_{GW}=0.67$ Gyr
(see also Valiante \& al. 2009).

An important parameter, introduced in the calculation of the rate, is
the constant $A_{Ia}$ that represents the fraction of systems which
are able to originate a SNe Ia explosion. The value of this constant
is calculated a posteriori and it is chosen so as to ensure that the
predicted present day SN Ia rate is reproduced. The assumed Type Ia SN
rates is given by Cappellaro \& al. (1999), that is $0.18 \pm 0.06
SNu$, where $1SNu=1SN/10^{10}L_{\odot B}/century$.

It is necessary to consider that an elliptical galaxy of initial
luminous mass of $10^{11}M_\odot$ at the present time has a lower
luminous mass, because of the presence of the galactic winds.  Then
the value of the constant $A_{Ia}$, that we have computed, is relative
to a galaxy with a stellar mass of $\sim 3.5 \cdot 10^{10}M_{\odot}$,
that is the final mass of the galaxy. As demonstrated by Valiante \&
al. (2009), to obtain the present time Type Ia SN rate, in units of
$SNe \cdot century^{-1}$, for an elliptical with a stellar mass of
$3.5\cdot 10^{10}M_{\odot}$, one should multiply the Cappellaro \&
al. rate by the blue luminosity predicted for a such galaxy, obtaining
a Type Ia SN rate of $0.072 SNe century^{-1}$ (see Valiante \&
al. 2009).  So we have considered this value for the SN Ia rate as
reference.

\subsection{Properties of different Type Ia SN rates}

\begin{figure}
\begin{center}
\includegraphics[scale=.40]{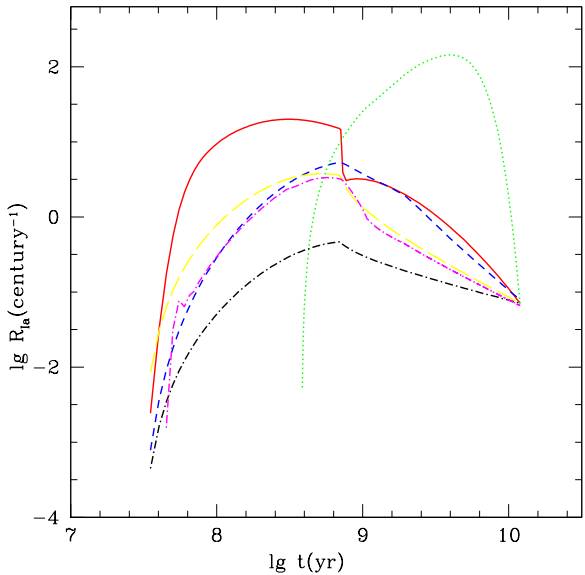} 
 \caption{The Type Ia SN rate, as a function of time, for an  elliptical galaxy of $10^{11}M_\odot$. Symbols are like in Figure 1.}
\label{rz} 
\end{center}
\end{figure}

Figure \ref{rz} represents the predicted Type Ia SN rates as functions
of redshift for the different DTDs. From this Figure it is possible
check that all the curves have a similar trend, except for the rate of
S04. This difference in the behaviour of the rate of S04 is due to the
fact that this DTD does not contain any prompt SNe Ia and then the
peak of this curve is shifted at longer times (several Gyrs). Also the
curve for the rate of Pritchet has a particular trend; in fact, it is
possible to verify that the peak of this it is much lower compared to
all the other curves.  The trend of the rate of Totani \& al. (2008)
is very similar to the rate of Greggio (2005) for wide binaries, in
agreement with the conclusions of Totani \& al. (2008).

In Table \ref{cdtd} we report the values derived from eq. (5) of the
various values of the parameter $A_{Ia}$, that we considered constant
in time, used in the different DTD models in order to obtain a present
time SN Ia rate in agreement with observations for a typical
elliptical. However, these values for $A_{Ia}$ are only indicative
since they change if the history of star formation changes.

\begin{table}
\begin{center}
\begin{tabular}{|c|c|}
\hline
\hline
Model & $A_{Ia}$ \\
\hline
\bfseries Mannucci \& al. (2006) & 0.0053 \\
\hline
\bfseries Matteucci \& Recchi (2001) & 0.015\\
\hline
\bfseries Pritchet \& al. (2008) & 0.00025\\
\hline
\bfseries Totani \& al. (2008) & 0.0013 \\
\hline
\bfseries Strolger \& al. (2004) & 0.028\\
\hline
\bfseries Greggio (2005) & 0.0002\\
\hline
\hline
\end{tabular}
\end{center}
\caption{Values of the normalization constant $cost_{dtd}$ and $A_{Ia}$ used in the different models. The models refer to the same SF history, the one of Figure 2, but to different DTDs.}
\label{cdtd}
\end{table}

\section{A more self-consistent computation of the Type Ia SNe rate in a typical elliptical}

The above analysis has thus allowed us to emphasize how the DTD
function may affect the calculation of the SN Ia rate. With the
purpose of having a more precise estimate of the SN Ia rate, we have
recalculated the various Type Ia SN rates by adopting the SFR
predicted by the chemical evolution model for the various DTDs. In
fact, according to the various DTDs, the galactic wind occurs at
different times, thus changing the history of SF by truncating it at
different times.  We recall that a galactic wind occurs when the
energy injected into the ISM by SNe equates the gas binding energy.
This approach is clearly more self-consistent relative to previous one
and it allows us to compute in detail the Fe production in
ellipticals, as we will see in Section 5. This fact was not considered
in Valiante \& al. (2009).

Due to the high similarity of Totani \& al. (2008) DTD and that of Greggio (2005) for wide binaries, we will present since now on only the model with the Totani et al. (2008) DTD ($\propto t^{-1}$).

\subsection{Star Formation Rates}
The histories of star formation dictated by the various DTDs are shown
in Fig. 4: one can immediately see how the only SFR that predicts the
occurrence of the galactic wind before $\sim 0.67 Gyr$, that is the
value relative to the previous SFR relative to the DTD of MR01 (SD
scenario), is the SFR relative to the bimodal DTD of Mannucci \&
al. (2006). In this model, in fact, the birth of a galactic wind
occurs at $\sim 0.46 Gyr$, due to the larger number of prompt Type Ia
SNe in this DTD.  All the other SFRs are lasting for a longer time. On
the other hand, the system in which the galactic wind occurs latest is
the one related to Pritchet \& al. (2008) DTD; in this system, in
fact, the SFR vanishes at $\sim 2.64 Gyr$, because of the absence of
prompt Type Ia SNe. In this case, a large fraction of the Fe produced
by Type Ia SNe remains trapped into stars. In the SFR with Totani \&
al. (2008) DTD the galactic wind occurs at $ \sim 1.6 Gyr$, while the
galactic wind time obtained with S04 is $\sim 1.75 Gyr$.

The values predicted by the different models for the occurrence of the
galactic wind, the total number of SNe Ia exploded during the Hubble
time, and the total mass of Fe they produced, are shown in Table
2. The total mass of Fe produced by SNe Ia has been computed by
assuming that each SNIa produces $0.6M_{\odot}$ of Fe (Iwamoto \&
al. 1999).

\begin{table}
\begin{center}
\begin{tabular}{|c|c|c|c|}
\hline
\hline
Model & $t_{GW}$ & SNIa & $M_{Fe}$ \\
&        (Gyr)&& (M$_{\odot}$)\\
\hline
\bfseries Mannucci \& al. (2006) & 0.46 & $2.12 \cdot10^8$ & $1.27\cdot10^8$ \\
\hline
\bfseries Matteucci \& Recchi (2001) & 0.67& $1.47 \cdot 10^8$ & $8.80 \cdot 10^7$ \\
\hline
\bfseries Pritchet \& al. (2008) & 2.64 &  $1.5 \cdot10^7$ & $9.00\cdot10^6$ \\
\hline
\bfseries Totani \& al. (2008) & 1.60 & $4.74 \cdot10^7$ & $2.84\cdot10^7$ \\
\hline
\bfseries Strolger \& al. (2004) &  1.75 & $1.68 \cdot10^9$ & $1.00\cdot10^9$ \\
\hline
\hline
\end{tabular}
\end{center}
\caption{Values of the $t_{GW}$, number of SNe Ia, e total mass of Fe produced by SNeIa ( $M_{Fe}(M_{\odot})$) in the different models for a typical elliptical, as described in the text.} 
\label{tgw}
\end{table}

\begin{figure}
\begin{center}
\includegraphics[scale=.40]{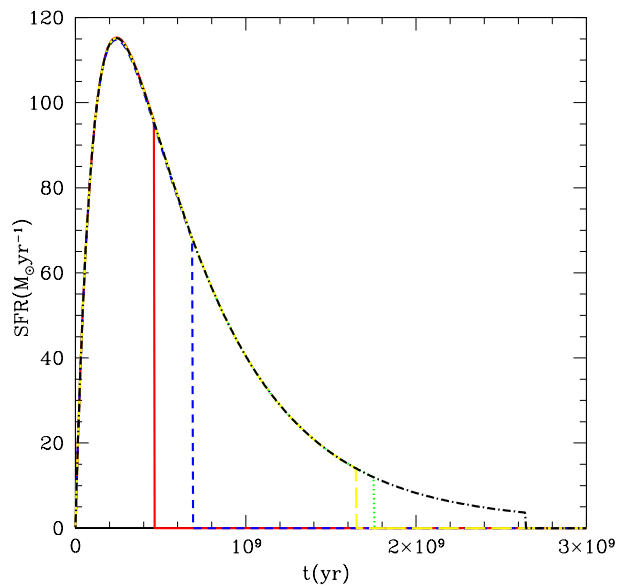}
\caption{Illustration of the SFR, as a function of time, for an  elliptical galaxy of $10^{11}M_\odot$. As one can see, the shape of the SFR is the same for all the cases but the SFR is truncated at different times in the different cases. 
Symbols are like in Figure 1.}
\label{SFRt} 
\end{center}
\end{figure}

\subsection{Type Ia SN rates}

The results of the Type Ia SN rates that we have obtained by
exploiting the SFRs determined by the different DTDs are reported in
Fig. 5.   Clearly the Type Ia SN rates so derived present
  differences relative to those of Figure 3 and they are due to the
  different duration of the star formation in the various
  cases. Different DTDs influence differently the onset of the
  galactic wind, as we have discussed before, and therefore the
  behaviour of the Type Ia SN rate. Moreover, the value of the
  constant $A_{Ia}$ undergoes some changes.  Varying the SFR, in fact,
  it is again necessary to ensure that the rate calculated is able to
  reproduce the current value by introducing the appropriate value of
  this constant. For the computation of the amount of Fe
  ejected into the ICM, that we will discuss in the next section, is
  the time of the occurrence of the galactic wind which influences the
  amount of Fe remaining trapped in stars versus the amount of Fe
  which can be ejected into the ICM (namely, all the Fe produced after
  SF has stopped).

\begin{figure}
\begin{center}
\includegraphics[scale=.40]{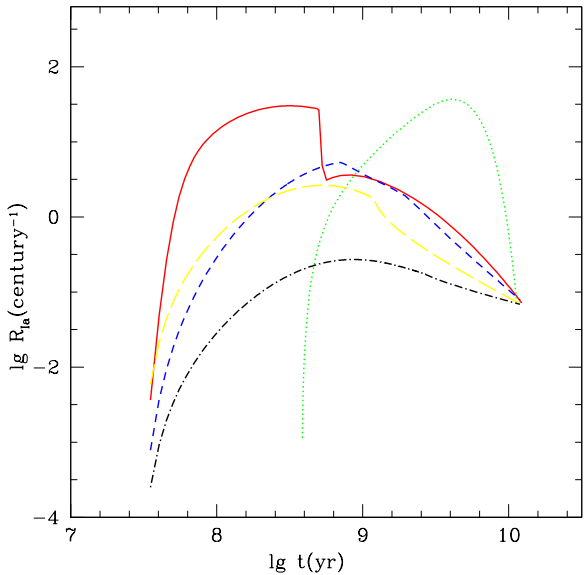}
\caption{Illustration of the different Type Ia SN rates, as a function of time, for an  elliptical galaxy of $10^{11}M_\odot$. Symbols are like in Figure 1.}
\label{Ras} 
\end{center}
\end{figure}

\section{Chemical enrichment of the ICM with different DTDs}
As a test for the different DTDs, we computed the expected Fe
enrichment in galaxy clusters. In particular, we considered only two
clusters: Virgo and Coma, taken as prototypes of a poor and a rich
cluster respectively. The method we adopted is similar to that of
Matteucci \& Vettolani (1988), Matteucci \& Gibson (1995), and Pipino
\& al. (2002), where the masses produced in the form of Fe and total
gas by ellipticals of different masses were integrated over the mass
function of the clusters. We remind that the main results of these
previous papers were that the Fe mass in clusters is easily reproduced
whereas the total gas mass ejected by galaxies is far lower than
observed. The logic conclusion from that was that most of the gas in
clusters is primordial. We find the same result here. However, here we
are interested in comparing the total Fe masses produced by
ellipticals evolving with different DTDs. We have considered only four
DTDs: i) the MR01 DTD, ii) the Mannucci \& al. DTD, iii) the Totani \&
al. DTD which is similar to the DTD relative to the DD wide scenario,
and iv) the S04 DTD. We did not include the Pritchett DTD since it is
clear from Table 2 that it predicts a very low number of SNe Ia and
therefore of Fe to be ejected into the intracluster medium (ICM). In
Table 3 we show the masses of Fe and gas, ejected into the ICM by
ellipticals of baryonic masses $10^{10}, 10^{11}$, and
$10^{12}M_{\odot}$.  The chemical evolution model adopted is the one
described in Section 2: the assumed efficiencies of star formation are
$3 Gyr^{-1}$, $10 Gyr^{-1}$, and $20Gyr^{-1}$ for $10^{10}$,
$10^{11}$, and $10^{12}M_{\odot}$, respectively (see Pipino \&
Matteucci, 2004).
 
\begin{table}
\caption{Masses ejected into the ICM by ellipticals in clusters}
\begin{center}
\begin{tabular}{c|ccc}
  \hline
\hline
\\
DTD   & M$_G$ & M$_{gas}(M_{\odot})$ & M$_{Fe}(M_{\odot})$ \\
\\
\hline
\\
Mannucci    & $10^{10}$ &  $2.26 \times 10^{9}$     &  $8.24 \times 10^{6}$  \\
\\

    & $10^{11}$ &  $1.08 \times 10^{10}$     &  $3.69 \times 10^{7}$  \\
\\
   & $10^{12}$ &  $8.32 \times 10^{10}$     &  $3.30 \times 10^{8}$    \\
\\
\hline
\\
Totani    & $10^{10}$ &  $9.66 \times 10^{8}$     &  $1.94 \times 10^{6}$   \\
\\
  & $10^{11}$ &  $7.13 \times 10^{9}$     &  $1.39 \times 10^{7}$      \\
\\
 & $10^{12}$ &  $6.00 \times 10^{10}$     &  $1.22 \times 10^{8}$      \\
\\
\hline
\\
Matteucci \& Recchi    & $10^{10}$ &  $1.12  \times 10^{9}$     &  $2.40 \times 10^{7}$  \\
\\
  & $10^{11}$ &  $1.04 \times 10^{10}$     &  $3.25 \times 10^{8}$   \\
\\
 & $10^{12}$ &  $8.62 \times 10^{10}$     &  $3.05 \times 10^{9}$     \\
\\
\hline
\\
Strolger \& al. &  $10^{10}$ &  $1.37 \times 10^{9}$& $4.31 \times 10^{7}$\\
\\
 &$10^{11}$ & $8.44 \times 10^{9}$ & $4.61 \times 10^{8}$ \\
\\
 & $10^{12}$ & $7.54 \times 10^{10}$ & $4.79 \times 10^{9}$
\\
\\
\hline
\end{tabular}
\end{center}

\end{table}

In the first column of Table 3 is indicated the DTD adopted in the chemical evolution model, in column 2 the initial baryonic mass of the galaxy, in column 3 the ejected total gas (H, He plus heavier elements), in column 4 the total ejected Fe mass.

Clearly, the Fe masses ejected into ICM in the different cases are
quite different, because the time at which the galactic wind occurs
and SF stops is different in different galaxies.  In Tables 4 and 5 we
show the total mass of Fe and gas ejected into the ICM after
integrating the single galactic contributions over the cluster mass
function for Virgo and Coma, respectively. In the same Tables are
shown the observed values for $M_{gas}$ and $M_{Fe}$.  For Virgo we
assumed the following values of the parameters necessary for the
integration (see Matteucci \& Vettolani, 1988 for details): $M/L=10$
(typical mass to light ratio of ellipticals), $\alpha=-1.25$ (slope of
the Schechter (1976) function), $f=0.43$ (fraction of ellipticals in
the cluster), $n^*=20$ (cluster richness), and $M_*=-22$ (absolute
magnitude of the galaxy at the break of the luminosity function). For
Coma we have adopted: $M/L=10$, $\alpha=-1.25$, $f=0.82$, $n^*=115$,
$M_*=-22$. The assumed Hubble constant is 67 $kmsec^{-1}Mpc^{-1}$.

\begin{table}
\caption{Virgo: integrated gas and Fe masses}
\scriptsize 
\label{models}
\begin{center}
\begin{tabular}{c|cc}
  \hline
\hline
\\
DTD   &  M$_{gas}(M_{\odot})$& M$_{Fe}(M_{\odot})$\\
\\
\hline
\\
Mannucci    &   $6.16 \times 10^{11}$     &  $2.33 \times 10^{9}$  \\
\\
\hline
\\
Totani    &  $3.42 \times 10^{11}$     &  $6.90 \times 10^{8}$   \\
\\
\hline
\\
Matteucci \& Recchi    &  $4.45 \times 10^{11}$    &  $1.29 \times 10^{10}$  \\

\\
\hline\\

Strolger \& al.   & $4.55 \times10^{11}$  &  $2.12 \times 10^{10}$\\
\\
\hline\\

Observed values        &  $2 \cdot 10^{13}$         &  $1.6 \cdot 10^{10}$ \\    \\       
\hline

\end{tabular}
\end{center}
\end{table}

\begin{table}
\caption{Coma: integrated gas and Fe masses}
\scriptsize 
\label{models}
\begin{center}
\begin{tabular}{c|cc}
  \hline
\hline
\\
DTD   &  M$_{gas}(M_{\odot})$& M$_{Fe}(M_{\odot})$\\
\\
\hline
\\
Mannucci    &     $6.75 \times 10^{12}$     &  $2.56 \times 10^{10}$ \\
\\

\hline
\\
Totani    &  $3.74 \times 10^{12}$     &  $7.56 \times 10^{9}$   \\
\\

\hline
\\

Matteucci \& Recchi    & $4.88 \times 10^{12}$   &  $1.41 \times 10^{11}$  \\
\\
\hline \\
Strolger \& al.        & $4.99 \times 10^{12}$&   $2.33 \times 10^{11}$\\
\\

\hline\\
Observed values    &  $(4.4 \pm 1.2) \cdot 10^{14}$ &  $3.1 \cdot 10^{11}$   \\
\\
\hline
\end{tabular}
\end{center}
\end{table}

From Tables 4 and 5 we can see that in the Mannucci \& al. DTD case,
the Fe ejected into the ICM is less than in the SD case (MR01 DTD):
this is because the number of prompt Type Ia SNe, in this DTD, is
quite large and 50\% of all SNe Ia explode inside 100 Myr, before the
galactic wind. As a consequence, most of the Fe produced by these SNe
will be incorporated into stars at variance with what happens for the
DTDs of MR01 and Totani, where the fraction of prompt SNe is much
lower. However, in the Totani DTD($\propto t^{-1}$) , although the
number of prompt SNe is lower, the total number of SNe Ia is also
lower than in the other cases. So, the smallest amount of Fe ejected
into the ICM is the one from Totani's DTD and is not enough to explain
the Fe observed in both clusters.  On the other hand, the integrated
values for Fe in the MR01 DTD and in the S04 DTD produce Fe masses in
agreement with observations. It is worth noting that the S04 DTD
predicts the smallest masses of total gas ejected into the ICM; this
is due to the fact that the galactic winds in ellipticals occur latest
with this DTD. Therefore, the amount of residual gas at the time of
the wind is lower than in the other cases. If we compute the Fe
abundance in the ICM by dividing the total Fe mass predicted for each
cluster by the observed mass of gas we obtain, for the MR01 DTD,
values in very good agreement with observations ($\sim 0.3Fe_{\odot}$,
Renzini, 2004): in particular, for Virgo we obtain: $Fe_{Virgo}\sim
0.4-0.5 Fe_{\odot}$ and for Coma $Fe_{Coma} \sim 0.2-0.3 Fe_{\odot}$,
having assumed the solar Fe abundance by Asplund \& al. (2009)
($Fe_{\odot} \sim 1.34 \cdot 10^{-3}$ by mass).

\section{The cosmic Type Ia SN Rate}

Another goal that we have set is to test the six different DTDs,
previously analyzed, in the computation of the cosmic Type Ia SN   
rate. The determination of the cosmic Type Ia SN rate, as well as the
calculation of the Type Ia SN rate, is obtained as the convolution
between the assumed cosmic star formation rate (CSFR) and the DTD
function. The CSFR is the star formation rate in a unitary comoving
volume of the universe and it is expressed in units of $M_{\odot}
yr^{-1} Mpc^{-3}$. For each of the DTD, that we will use, we will
adopt five different cosmic star formation rates, that is: i) Cole \&
al. (2001) modified (see later) CSFR; ii) Madau, Della Valle \&
Panagia (1998) MDP01 and MDP02 CSFRs; iii) S04 CSFR; iv) Grieco \&
al. (2012) CSFR.  The assumed cosmology, as stated before, is the
Lambda Cold Dark Matter ($\Lambda$CDM) cosmology.

\subsection{The cosmic star formation rate}

The physical meaning of the CSFR is of cumulative SFR owing to
galaxies of different morphological types present in a unitary
comoving volume of the Universe. The CSFR is not a directly measurable
quantity and it can be computed only from the measurement of the
luminosity density in different wavebands, which are then transformed
into star formation rate by a suitable calibration. At high redshift,
in fact, it is difficult to distinguish galaxy morphology but it is
only possible to trace the luminosity density of galaxies.

\subsubsection{Cole \& al.(2001) CSFR}
The cosmic star formation rate density proposed by Cole \& al. (2001) has the form:

\begin{equation}
\dot{\rho}_{Cole}=\frac{a+bz}{1+(\frac{z}{c})^d} 
\end{equation}
where $(a,b,c,d)=(0.0,0.0798,1.658,3.105)$ and after the correction
for the dust extinction are $(a,b,c,d)=(0.0166,0.1848,1.9474,2.6316)$.
This parametric fit of the CSFR was obtained thanks to the analysis of
the data of Two Micron All Sky Survey (2MASS) Extended Source Catalog
and the 2dF Galaxy Redshift Survey. Cole \& al. (2001) used these data
to estimate the galaxy luminosity function and to infer the total mass
fraction in stars.  In the computation of this CSFR, Cole \&
al. (2001) have assumed that no mass goes into forming brown dwarfs
and they have multiplied the star formation rate by $(1-R)$ where $R$
is the recycled fraction, as defined in the Simple Model of chemical
evolution (Tinsley 1980), they have obtained an estimate of the mass
locked up in stars.  Here we have done the best fit of all the data on
CSFR previously used by Cole \& al. (2001) plus some new ones (see
Vincoletto \& al. 2012 for the data and references).The obtained best
fit is very similar to that of Cole \& al. (2001) and it is shown in
Figure 6 together with the other CSFRs adopted in this paper. We will
refer to it as the modified Cole CSFR.  For the modified Cole CSFR we
have adopted the following parameters: a=0.00904, b=0.1122, c=3.325,
d=4.143.

\begin{figure}
\begin{center}
\includegraphics[scale=.40]{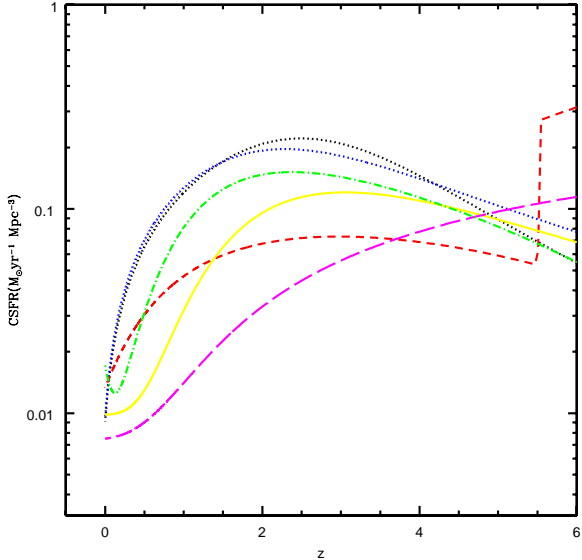}
\caption{Illustration of the various cosmic star formation
  history. The dotted blue line is the CSFR of Cole \& al. (2001)
  whereas the black dotted line is the modified Cole CSFR (see text);
  the solid yellow line is the CSFR of MDP01; the long dashed magenta
  line is the CSFR of MDP02; the dashed-dotted green line is the CSFR
  of S04 (2004); the short dashed red line is the CSFR of Grieco \&
  al. (2012).}
\label{CSFRs} 
\end{center}
\end{figure}

\begin{figure}
\begin{center}
\includegraphics[scale=.40]{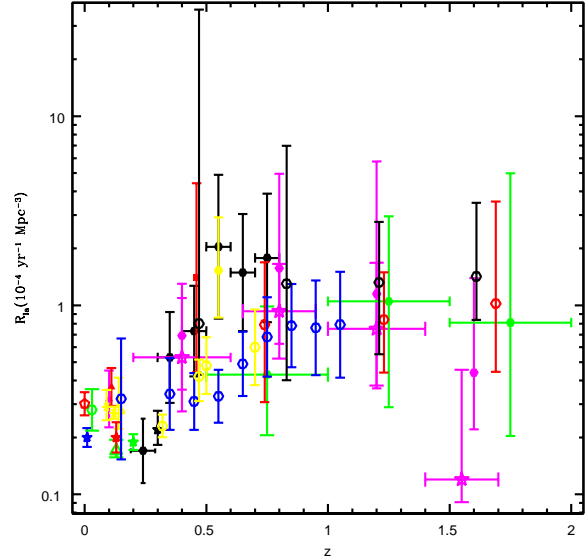}
\caption{Illustration of the observational data: Barris \& Tonry
  (2006) (black filled hexagons), Dahlen et al. (2004) (magenta filled
  hexagons), Kuznetsova et al. (2008) ( magenta open stars), Poznanski
  et al. (2007) (green filled hexagons), Cappellaro et al. (1999)
  (blue open stars), Hardin et al. (2000) (yellow open triangles),
  Blanc et al. (2004) (green open triangles), Mannucci et al. (2005)
  (green open hexagons), Magdwick et al. (2003) (magenta filled
  triangles), Strolger et al. (2003) (red filled hexagons), Neil et
  al. (2007) (yellow open hexagons), Horesh et al. (2008) (green open
  stars), Botticella et al. (2008) (black open stars), Tonry et
  al. (2003) (red filled squares). Neill et al. (2006) (blue filled
  pentagons), Pain et al. (2002) (yellow filled hexagons), Graur et
  al. (2011) (red open hexagons), Rodney \& Tonry (2010) (blue open
  hexagons), Li et al. (2010) (red open pentagons), Dilday et
  al. (2010) (yellow filled stars), Dahlen et al. (2008) (black open
  hexagons).}
\label{data}
\end{center}
\end{figure}

\subsubsection{MDP (1998) CSFR}

The two cosmic SFR models MDP01 and MDP02 computed in Madau \& al.(1998) are:

\begin{eqnarray}
\dot{\rho}_{MDP01}(t)=a_1[t_9^{a_2}e^{-{\frac{t_9}{a_3}}}+a_4(1-e^{-{\frac{t_9}{a_3}}})] \\
\dot{\rho}_{MDP02}(t)=a_1e^{-{\frac{t_9}{a_6}}}+a_4(1-e^{-{\frac{t_9}{a_3}}})+a_5t_9^{a_2}e^-{\frac{t_9}{a_3}}
\end{eqnarray}

where: 
\begin{description}
  \item $t_9= 13(1+z)^{-(\frac{3}{2})}$ is the Hubble time at redshift z;
  \item $a_1= 0.049$ \,\, in\,\, MDP01, \,\,  $a_1= 0.336$\,\,  in\,\, MDP02;
  \item $a_2=5$\,\, in\,\, both\,\, cases;
  \item $a_3=0.64$\,\, in \,\, both \,\, cases;
  \item $a_4= 0.2$\,\, in\,\, MDP01, \,\,  $a_4= 0.0074$\,\, in\,\, MDP02;
  \item $a_5=0.00197$;
  \item $a_6=1.6$
\end{description}

\subsubsection{Strolger \& al. (2004) CSFR}

This CSFR is given by:

\begin{equation}
\dot{\rho}_{S04}(t)=a_1(t^{a_2}e^{-{\frac{t}{a_3}}}+{a_4}e^{\frac{d(t-t_0)}{a_3}})
\end{equation}

where: 
\begin{description}
  \item $t_0=13.47Gyr$ is the age of the Universe;
  \item $a_1= 0.182$;
  \item $a_2=1.260$;
  \item $a_3=1.865$;
  \item $a_4=0.071$; 
\end{description}
These parameters are determined by fitting the collection of measurements of the CSFR of Giavalisco \& al. (2004).
This CSFR is a model based on a modified version of the parametric form suggested by Madau et al. (1998), taking into account dust extinction.

\begin{center}
\begin{figure*}
\begin{tabular}{cc}
\includegraphics[scale=.4]{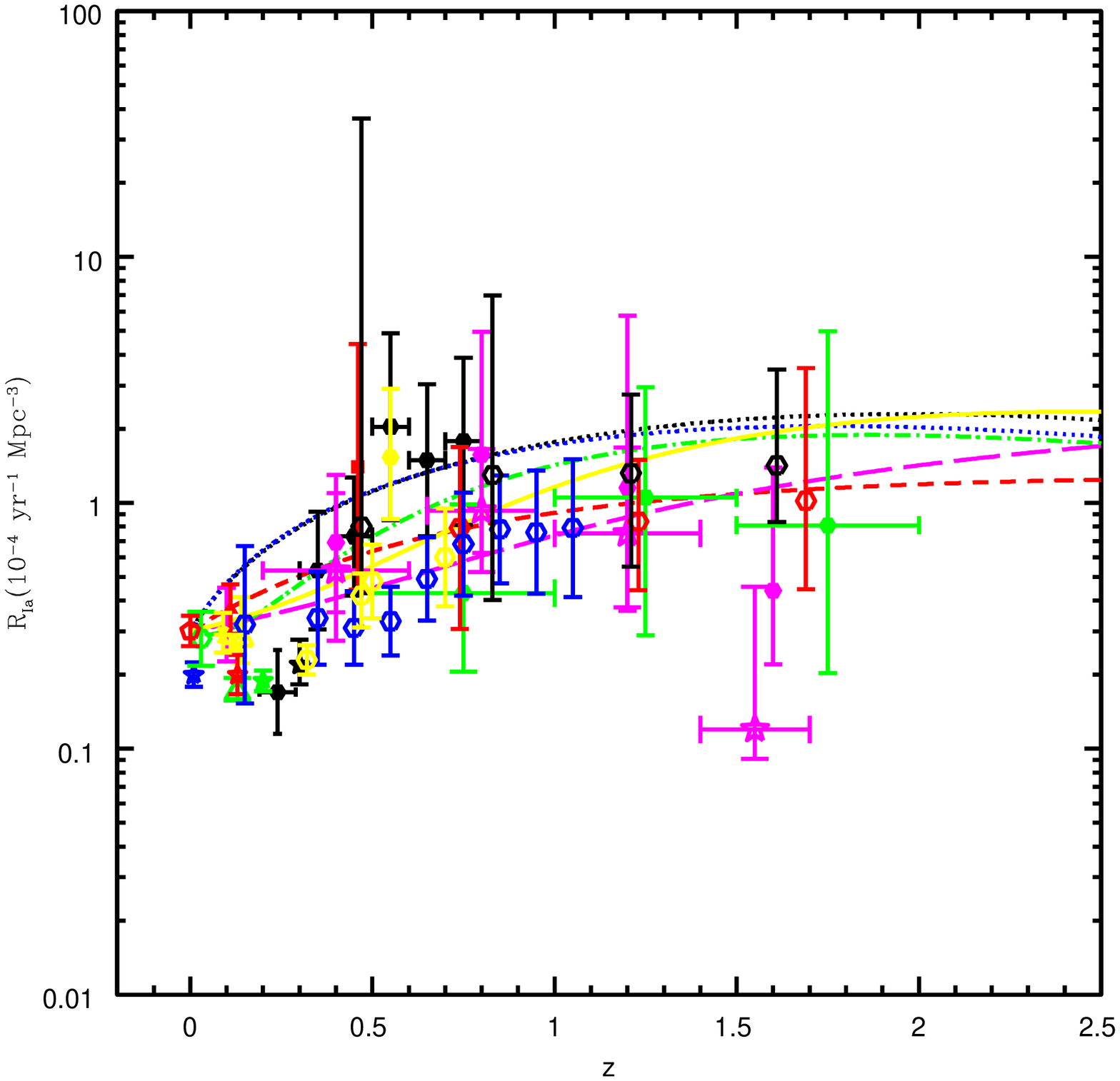} &
\includegraphics[scale=.4]{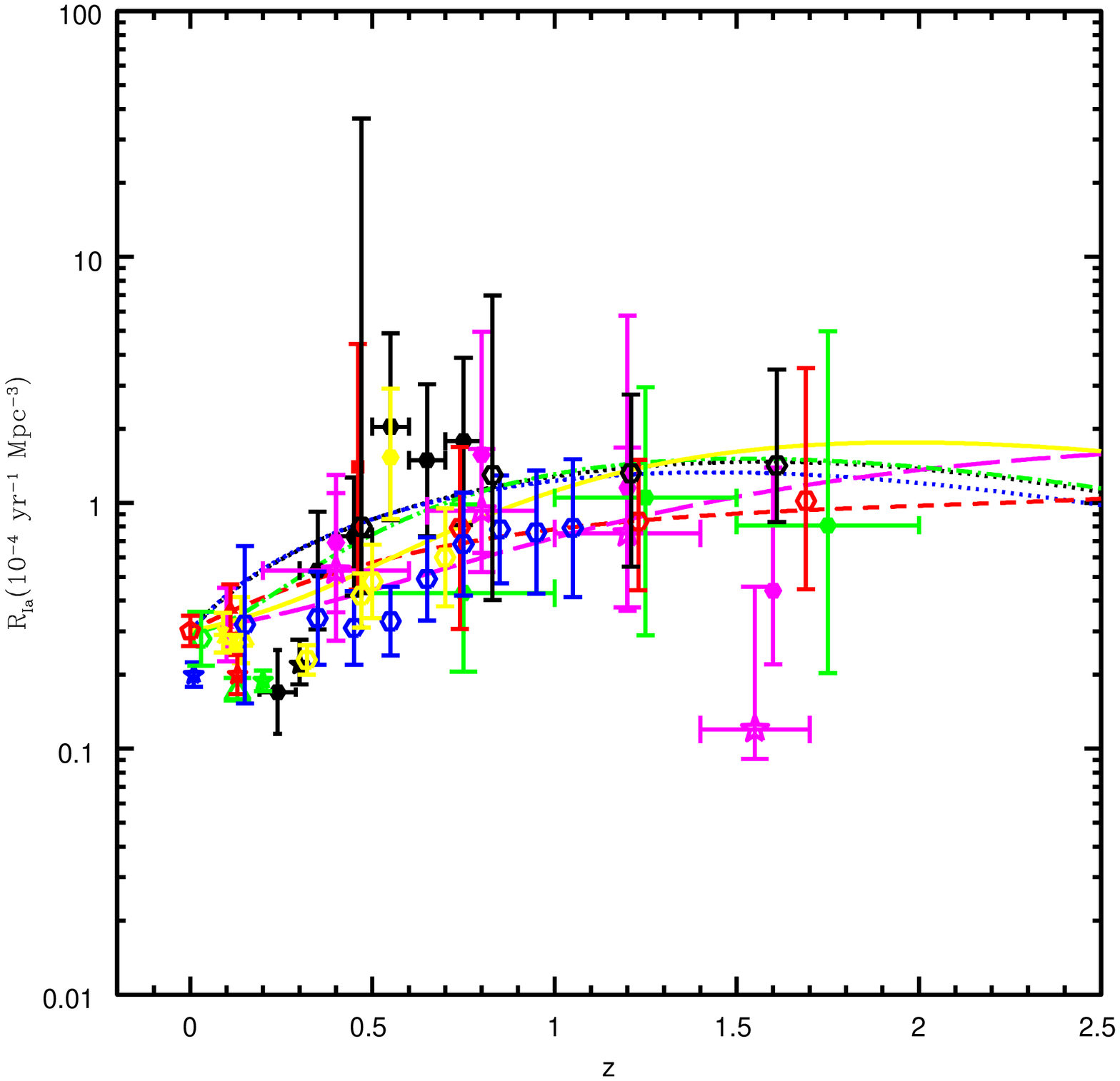} \\
\includegraphics[scale=.4]{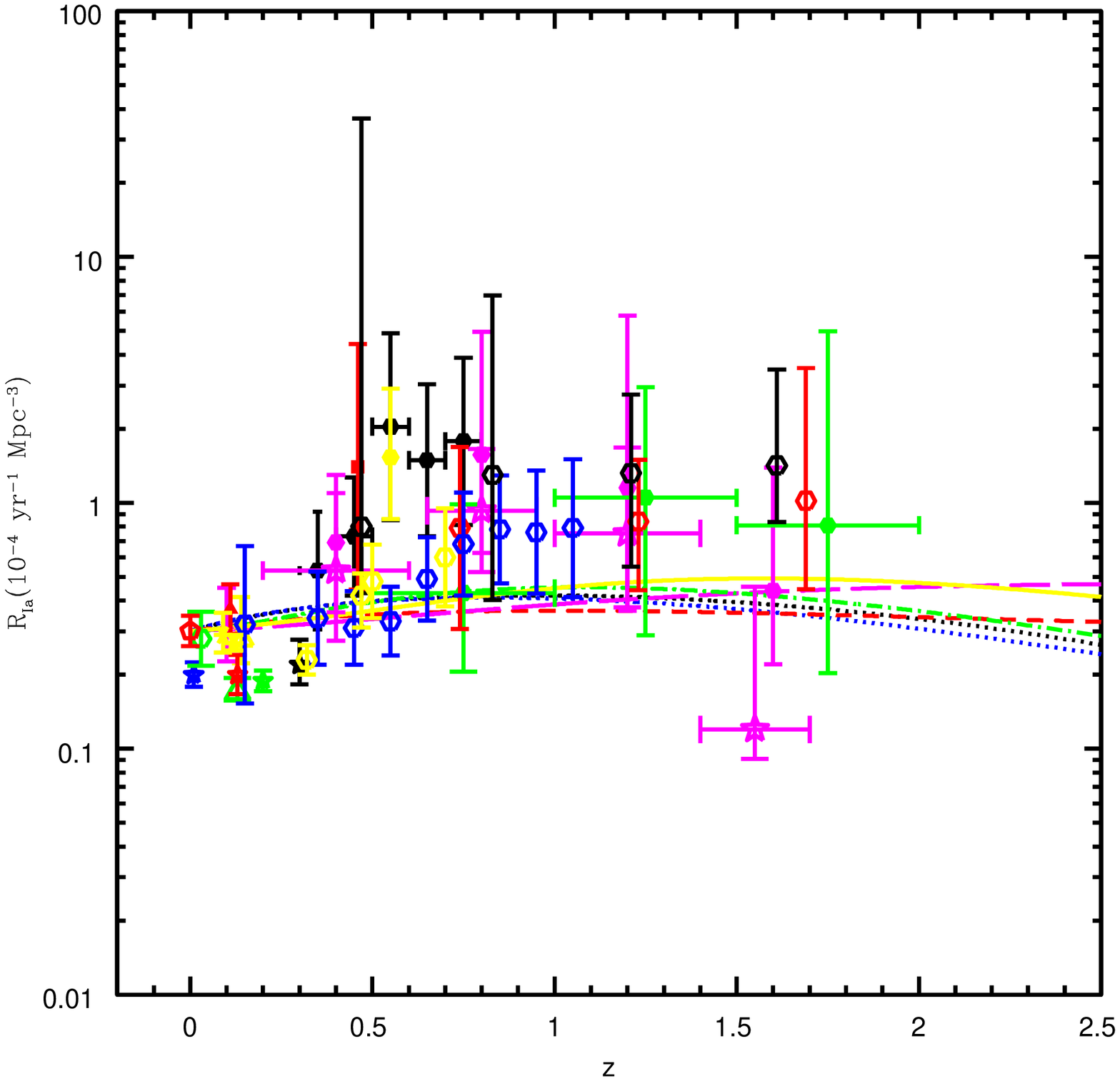} &
\includegraphics[scale=.4]{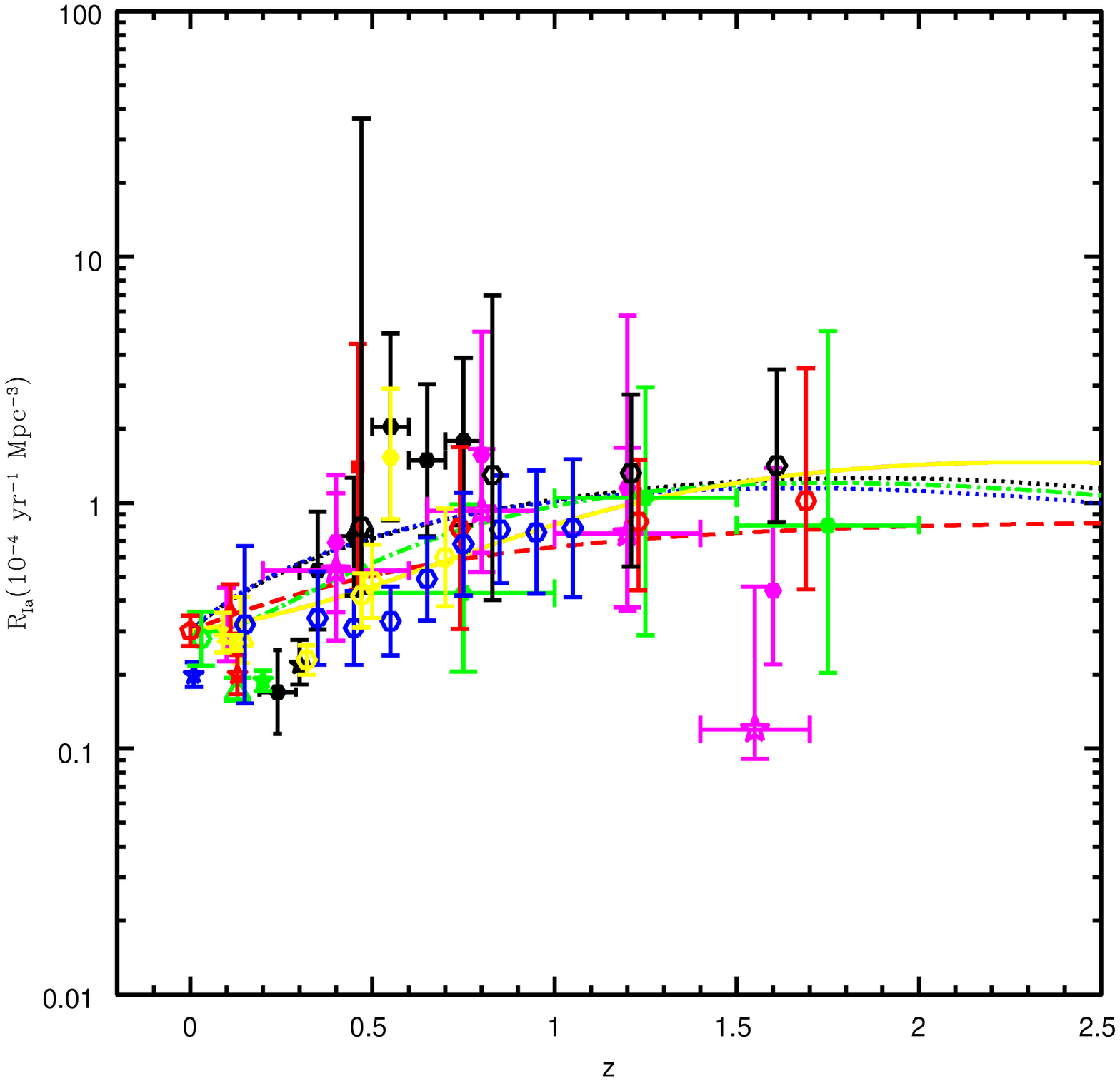} \\
\end{tabular}
\caption{Comparison between model and data. The two top panels are
  related to the MVP06 DTD, one on the left, and the other to the MR01
  DTD plied to the 5 different CSFRs of Figure 6 (the symbols are the
  same as in Fig.7). In the last two panels there are the Pritchet \&
  al. DTD, on the left, and on the right the Totani \& al. DTD, also
  applied to the 5 CSFRs of Fig.6.}
\label{Rattot}
\end{figure*}
\end{center}

\subsubsection{Grieco \& al. (2012) CSFR}

To compute the CSFR Grieco \& al. (2012) have adopted the following relation:

\begin{equation}
\dot{\rho}_{Grieco}=\sum_{k}\psi_k(t)\cdot n_k^{\star}  (M_{\odot}\cdot yr^{-1} \cdot Mpc^{-3})
\end{equation}

where:
\begin{description}
\item $k$ identifies a particular galaxy type, that is elliptical, spiral or irregular;
\item $\psi_k(t)$  represents the history of star formation in each galaxy, tested by model of galactic chemical evolution;
\item $n_k^\star$ is the galaxy number density, expressed in units of $Mpc^{-3}$ for each morphological galaxy type, and it is been assumed to be constant and equal to the present time one; models assuming number density evolution can be found in Vincoletto \& al. (2012).
\end{description}

For the computation of this CSFR Grieco \& al. (2012) have assumed that all galaxies started forming stars at the same time, an oversimplified hypothesis but useful to understand the behaviour of the cosmic rates in extreme conditions.

\subsubsection{Trend of different CSFRs}

To make a comparison between all the CSFR that we analyzed we list all the 
trends in Figure \ref{CSFRs}.

Among the five different curves only two have a different behaviour,
that of Grieco \& al. (2012) and that of MDP02. The Grieco \&
al. (2012) CSFR, in fact, shows a quick rise at very high redshift
followed by a sharp decline and a following smooth decline until the
present time. The rising CSFR for $z>5$ depends on the fact that in
this model all the ellipticals are assumed to be present since the
beginning (no number density evolution). On the other hand, the MDP02
CSFR has only one continuous phase of slow growth and it mimics a
monolithic collapse scenario. In both CSFRs a Salpeter (1955) IMF is
assumed. These two curves therefore do not have the usual trend that
is common to the other three curves, that is an initial phase of
growth, the reach of a peak and finally a descending phase. All the
other models assume a hierarchical galaxy formation where the most
massive objects form later.

\subsection{Observational data}

The benefit of these new calculations is that, in this case, we can
compare the results of our models with a recent compilation of
observational data as function of redshift. The data that we have used
are reported in Appendix (Tables 6 and 7), where we indicate the
references, the redshift and the values of the cosmic SN Ia rate,
expressed in $10^{-4} Mpc^{-3} yr^{-1}$. The most recent ones are from
Graur \& al. (2011) and come from the SubaruDeep Field search.  The
data with their error bars are also shown in Figure \ref{data}.

\begin{center}
\begin{figure*}
\begin{tabular}{cc}
\includegraphics[scale=.40]{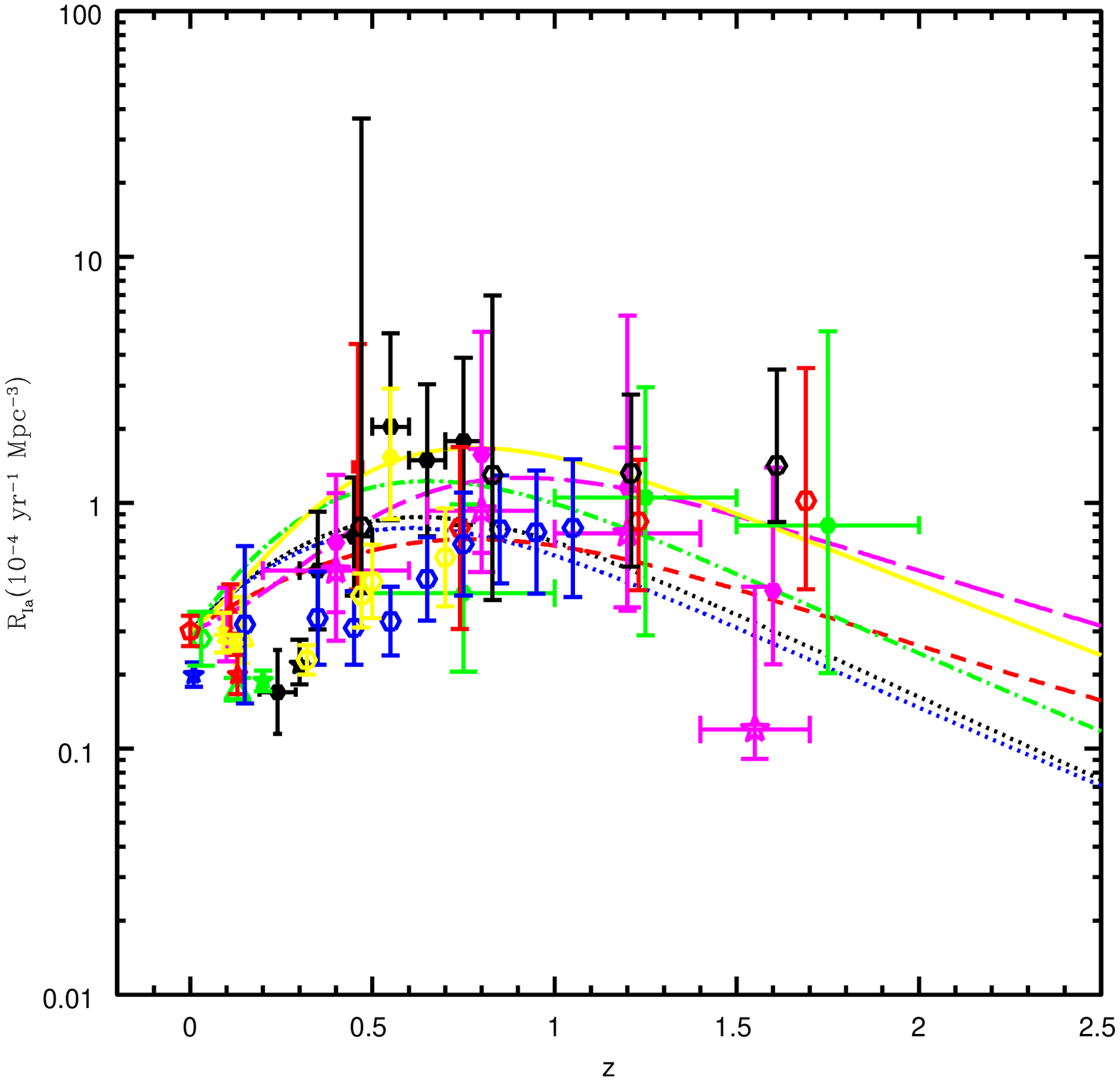} &
\includegraphics[scale=.40]{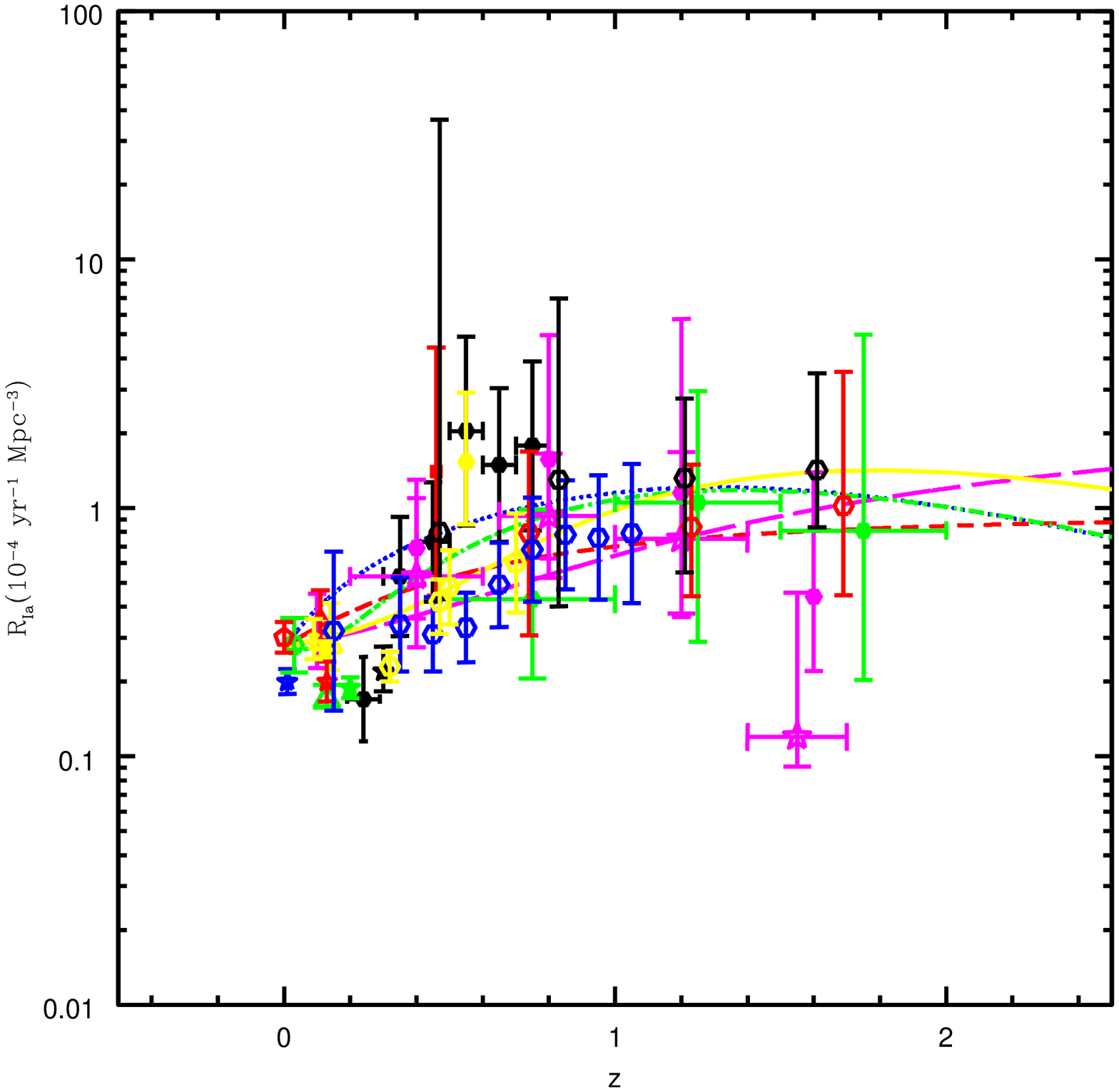} \\
\end{tabular}
\caption{Comparison between model and data. The two panels are related to the S04 DTD and to G05 DTD.}
\label{Rattot1}
\end{figure*}
\end{center}

\subsection{Computation of the cosmic Type Ia rate}

The computation of the cosmic Type Ia rate is, as already said,
similar to the computation of the Type Ia SN rate; so, also in this
case, it is necessary to introduce the constant $A_{Ia}$ to reproduce
the actual value of the cosmic SN Ia rate. The actual value 
  (z=0) of the cosmic Type Ia rate is $0.301
(10^{-4}Mpc^{-3}yr^{-1})$, as proposed by Li \& al. (2011).

\subsubsection{Properties of the different cosmic Type Ia SN rate}
The model results compared to the data are shown in Figures 8, 9 and
10.  In Figures 8 and 9, each model was obtained by the convolution
between a given DTD and five different CSFRs, shown in Figure 6. These
models are compared to the observed cosmic Type Ia SN rate.  From this
Figure we can see that the models that generally agree with most of
the data are unable to reproduce the data of Cappellaro \& al. (1999),
Blanc \& al. (2004). On the other hand, most of the models reproduce
the data of Graur \& al. (2011), Li \& al. (2011) and Rodney \& Tonry
(2010).  Moreover, it is worth noticing that the adopted data suggest
a minimum of the cosmic Type Ia SN rate at z $\sim$ 0.2--0.3 (see
Fig. 7). This is confirmed by a more quantitative analysis. The
average SNIa rate for the redshift bin $0<z\leq 0.1$ is 0.273, i.e. it
is lower than the present cosmic rate (0.301) determined by Li \&
al. (2011). The average SNIa rate further decreases in the redshift
bins $0.1<z\leq 0.2$ (0.244) and $0.2<z\leq 0.3$ (0.195). Then it
starts growing for $z>0.3$ (it is 0.448 in the redshift bin $0.3<z\leq
0.4$). None of the adopted combinations of CSFRs and DTDs is able to
reproduce this minimum. In order to reproduce it, we should either
assume a peculiar CSFR with a minimum at high z (but available data do
not corroborate it) or a DTD with a minimum at large delay times (but
from a theoretical point of view this is not easy to justify).

Looking at the Figures \ref{Rattot} and \ref{Rattot1} it is possible
to see how the models computed using the Pritchet \& al. (2008) DTD
are all too low compared to the data. The same effect seems to be
present for the S04 DTD. However, it is very difficult to select a
particular DTD and a particular CSFR as the best fit to the
data. First of all because of the large error bars present in the data
and the different way in which the errors have been computed in
different papers. Because of this, any conclusion concerning DTD and
CSFR should be taken with care.  Concerning the CSFR, we can see from
Figures \ref{Rattot} and \ref{Rattot1} that in all models, even in
those which are not able to fit very well the data, the curve relative
to the modified Cole \& al. (2001) CSFR predicts the best trend. With
the aim of studying this CSFR more in detail, we made a complementary
analysis in which we used only the CSFR of Cole \& al. (2001) and the
different DTDs. The result obtained is shown in Figure 10.  The same
considerations made for Figures 8 and 9 hold for this
Figure. Although, the modified Cole CSFR is probably the best CSFR
since it represents the fit to to the observed CSFR, it is still not
possible to extract the best DTD from this diagram. In fact, just
looking by eye one can conclude that the DTDs of Pritchett and S04 are
too low relative to the average of the data, whereas the DTDs of
Mannucci, MR01 and Totani are probably too high. For choosing the best
DTD, it is better to rely to chemical evolution results which adopt
elemental abundances measured with higher precision.  We adopted a
fitness test, as described in Calura \& al. (2010) (see their eq. 12),
to select the best combination DTD/CSFR and the results are shown in
Figure 11. From this figure it appears, that the modified Cole and the
Grieco \& al. (2012) CSFR are the best together with S04, MR01 and
Totani DTDs.  One can see that in the Figure 11, where the size of the
circle is inversely proportional to the fitness function, as defined
in Calura \& al. (2010).  The fact that MR01 and Totani 's DTDs are
the best is not surprising since they correspond to the two most
popular and tested Type Ia SN scenarios, the SD and the DD
ones. However, it is surprising that also the DTD of S04 is good since
it is very different from the other two and it does not reproduce the
chemical evolution of the solar vicinity (see Matteucci \&
al. 2009). The abundance patterns in the solar neighbourhood, in fact,
demand the presence of some prompt Type Ia SNe and a majority of tardy
Type Ia SNe. Therefore, no firm conclusions can be derived from the
cosmic Type Ia SN rate, since the uncertainties in the measured rates
prevent a reliable analysis relative to the DTDs.

\begin{figure}
\begin{center}
\includegraphics[scale=.40] {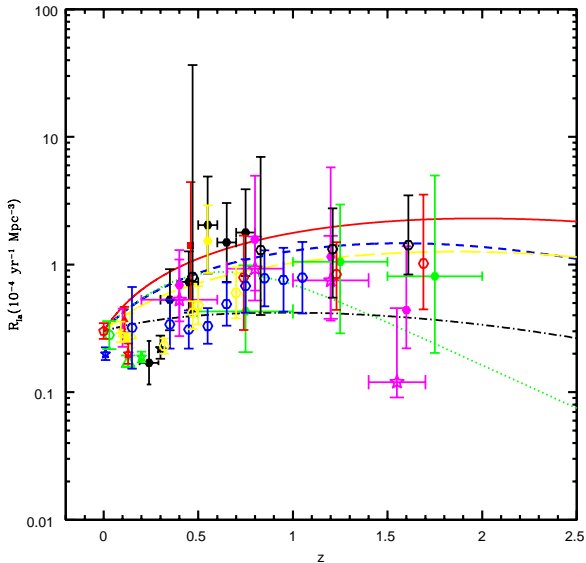}
\caption{Comparison between the model and the data, relative to the
  cosmic Type Ia SN rate computed using the modified Cole \& al (2001)
  CSFR and all the DTD. The solid red line is the cosmic Type Ia SN
  rate of MVP06; the short dashed blue line is the cosmic Type Ia SN
  rate of MR01; the dotted green line is the cosmic Type Ia SN rate of
  S04; the short dashed-dotted black line is the cosmic Type Ia SN
  rate of Pritchet \& al. (2008);  the long dashed yellow line is
    the cosmic Type Ia SN rate of Totani \& al. (2008). The cosmic
    Type Ia SN rate of G05 is not reported because is identical to the
    one of Totani \& al. (2008)}.
\label{col}
\end{center}
\end{figure}

\begin{figure}
\begin{center}
\includegraphics[angle=-90,scale=.40]{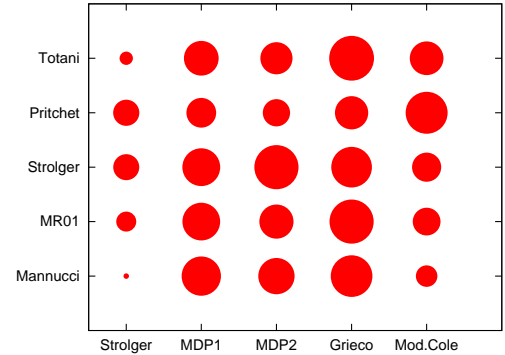}
\caption{Results of the fitness test for finding the best combination of CSFR and DTD. In the Y axis we report the various DTDs while in the X axis the various CSFRs. The size of the circle is inversely proportional to the calculated 
fitness function, as defined in Calura \& al. (2010), therefore better CSFR/DTD combinations appear as the largest circles.}
\label{pall}
\end{center}
\end{figure}

\section{Conclusions}
We have computed the Type Ia SN rates in elliptical galaxies of
various baryonic masses, under different assumptions about Type Ia SN
progenitors. To take into account the different SN Ia progenitors, we
adopted different delay time distribution functions (DTDs) describing
the distribution of the explosion times as a function of time. We
considered both theoretical and empirical DTD functions.  The
different DTDs have all being normalized to reproduce the present day
SNIa rate in typical ellipticals but they contains a different total
number of SNeIa and a different proportion of prompt (exploding in the
first 100 Myr since the beginning of star formation) and tardy SNe
Ia. The bimodal DTD contains $\sim50\%$ of prompt SNe, the SD DTD
(MR01) contains $\sim 13-15\%$, the DD DTD and the DTD $\propto
t^{-1}$ (Totani \& al. 2008) contain $\sim 10\%$, the DTD $\propto
t^{-0.5}$ (Pritchett \& al. 2008) contains $\sim 4\%$ and the DTD of
S04 contains zero.  Then, we have calculated the integrated Fe and gas
mass in two galaxy clusters (Coma and Virgo) by means of the models
for ellipticals including different DTDs.  Finally, we have studied
the cosmic Type Ia SN rate by adopting the same DTDs as for the
elliptical galaxies and by varying the assumed cosmic star formation
rate (CSFR). We considered CSFRs obtained either as bestfits of data
or theoretically, and containing different assumptions about galaxy
formation mechanisms.  We compared the predicted cosmic Type Ia SN
rates with the most recent compilation of data relative to the cosmic
Type Ia rate observed up to redshift $z=1.75$.  We have then compared
our model results with observations.  It is worth noting that the
results of this investigation should integrate those obtained by
Matteucci \& al. (2009). In that paper we tested the effects of
different DTDs on the chemical evolution of the Milky Way and we
obtained clear suggestions; in particular, we found that only the SD
and DD DTDs together with the bimodal DTD, but with less than 50\%
prompt Type Ia SNe, can fit the abundance patterns observed in the
stars of the Galaxy. Abundance measurements are very accurate these
days and certainly more accurate than cosmic SN rate ones. In
Matteucci et al. (2009) we also concluded that prompt Type Ia SNe are
necessary to reproduce the abundance data.

Our main conclusions can be summarized as follows:

\begin{itemize}

\item We have found that a different number of prompt Type Ia SNe
  affects substantially the time for the occurrence of a galactic
  wind, which then quenches star formation and gives rise to the
  passive evolutionary phase for such galaxies. Clearly, the time for
  a galactic wind influences the amount of Fe locked up in stars as
  well as that ejected into the ICM.

\item The best DTD in order to obtain the right amount of Fe in the
  ICM is the one relative to the single degenerate scenario in the
  formulation of Greggio \& Renzini (1983) and Matteucci \& Recchi
  (2001).  The DTD obtained by Totani \& al. (2008), which is similar
  to the DTD of  G05 DD wide scenario, does not produce enough Fe
  to be ejected into the ICM. This DTD, in fact, contains less SNe Ia
  than the  MR01 SD scenario and therefore the galactic wind
  occurs later. As a consequence of this, the Fe ejected into the ICM
  is less than in the MR01 SD scenario. The S04 DTD instead
  produces masses of Fe compatible with observations, but the amount
  of the total gas ejected is the lowest. This is due to the late
  galactic winds occurring in galaxies if this DTD is assumed.

\item The cosmic Type Ia SN rate is not a good tracer of the DTD nor
  of the CSFR.  In spite of the fact that we adopted the largest data
  set, the observed Type Ia SN rates are still quite uncertain and
  limited to a redshift z$\le 1.75$, and the error bars are quite
  large, especially at high z. We have performed a statistical
  analysis and found that the best DTDs seems to be those relative to
  the SD (MR01) and DD (G05) scenarios plus the DTD of
  S04. This last DTD was in fact deduced from a fit to the observed
  cosmic Type Ia SN rates. However, given the lack of prompt Type Ia
  SNe, this DTD does not reproduce correctly the chemical evolution of
  galaxies, as shown by Matteucci \& al. (2006; 2009). In particular,
  the lack of prompt Type Ia SNe produces a long plateau in the
  [$\alpha$/Fe] ratios in the Galaxy, at variance with
  observations. The opposite effect is obtained when the bimodal DTD
  is assumed with 50\% of prompt SNe. In this case [$\alpha$/Fe]
  ratios decrease too steeply with [Fe/H], at variance with
  observations.  The CSFRs adopted here are derived from both
  hierarchical and monolithic models of galaxy formation; in
  particular, different assumptions about the galaxy number density
  evolution underline these cases. In hierarchical models the galaxy
  number density varies with the cosmic time whereas in monolithic
  models the galaxy number density is assumed constant (pure
  luminosity evolution). Moreover, we took into consideration the fit
  of Cole \& al. (2001) of observational data and we also added other
  points and derived a revised best fit. From our analysis we suggest
  that the best ones to reproduce the cosmic Type Ia SN rate are that
  derived as a bestfit of observational data (Cole \& al. 2001 and
  this paper) and a theoretical CSFR derived from chemical evolution
  models (e.g. Grieco \& al. 2012)in the framework of pure luminosity
  evolution.

\item In summary, taking all of our results together we suggest that
  the best DTDs are those relative to the SD, DD and bimodal
  scenarios. All these DTDs contain prompt Type Ia SNe.  We also
  suggest that the ideal number of prompt Type Ia SNe should not
  exceed $15-20\%$ of the total SNe Ia.  However, these results
    are mainly supported by chemical evolution considerations rather
    than by the galactic and cosmic Type Ia SN rates. Therefore, the
    answer to the title of the paper is still rather negative. More
    precise data on SN Ia rates in galaxies and as a function of
    redshift will help in the future to find a more precise answer. 
\end{itemize}

\section*{Acknowledgments} {We thank E. Cappellaro for providing the data on the SN cosmic rates. F.M. acknowledges financial support from PRIN MIUR 2010-2011, project ``The Chemical and Dynamical Evolution of the Milky Way and Local Group Galaxies, prot. 2010LY5N2T.Finally, we thank an anonymous referee for his/her careful reading of the manuscript and important suggestions.}

\newpage
\begin{table*}
\begin{center}
\begin{tabular}{|c|c|c|}
\hline
\hline
Reference & z & SN rate($10^{-4} Mpc^{-3} yr^{-1}$) \\
\hline
Cappellaro \& al. (1999) & 0.01 & 0.20 $\pm$ 0.059 \\
\hline
Hardin \& al. (2000) & 0.14 & $0.22^{+0.17}_{-0.22}$ \\
\hline
Pain \& al. (2002) & 0.55 & $1.53^{+0.28}_{-0.25}$  \\
\hline
Magdwick \& al. (2003) & 0.10 & $0.32\pm 0.15$ \\
\hline
Strolger \& al. (2003) & 0.11 & $0.37\pm 0.10$  \\
\hline
Tonry \& al. (2003) & 0.46 & $1.4\pm 0.5$  \\
\hline
Blanc \& al. (2004) & 0.13 & $0.14^{+0.05}_{-0.035}$ \\
\hline
Dahlen \& al. (2004) & 0.4 & $0.69^{+0.34}_{-0.27}$  \\
\hline
Dahlen \& al. (2004) & 0.8 & $1.57^{+0.44}_{-0.25}$  \\
\hline
Dahlen \& al. (2004) & 1.2 & $1.15^{+0.47}_{-0.26}$ \\ 
\hline
Dahlen \& al. (2004) & 1.6 & $0.44^{+0.32}_{-0.25}$  \\
\hline
Mannucci \& al. (2005) & 0.03 & $0.28\pm 0.11$  \\
\hline
Barris \& Tonry (2006) & 0.25 & $0.17^{+0.17}_{-0.16}$  \\ 
\hline
Barris \& Tonry (2006) & 0.35 & $0.53\pm 0.24$  \\ 
\hline
Barris \& Tonry (2006) & 0.45 & $0.73\pm 0.24$  \\
\hline
Barris \& Tonry (2006) & 0.55 & $2.04\pm 0.38$  \\
\hline
Barris \& Tonry (2006) & 0.65 & $1.49\pm 0.31$  \\
\hline
Barris \& Tonry (2006) & 0.75 & $1.78\pm 0.34$  \\ 
\hline
Neill \& al. (2006) & 0.47 & $0.42^{+0.09}_{-0.13}$ \\
\hline
Neill \& al. (2007) & 0.32 & $0.23\pm 0.06$  \\
\hline
Neill \& al. (2007) & 0.50 & $0.48\pm 0.15$ \\
\hline
Neill \& al. (2007) & 0.7 & $0.60\pm 0.20$ \\
\hline
Poznanski \& al. (2007) & 0.75 & $0.43^{+0.36}_{-0.32}$ \\
\hline
Poznanski \& al. (2007) & 1.25 & $1.05^{+0.45}_{-0.56}$  \\
\hline
Poznanski \& al. (2007) & 1.75 & $0.81^{+0.79}_{-0.60}$ \\
\hline
\hline
\end{tabular}
\end{center}
\caption{A compilation of the observational cosmic Type Ia rates at different redshifts up to $z\sim1.75$.}
\label{dati} 
\end{table*}

\begin{table*}
\begin{center}
\begin{tabular}{|c|c|c|}
\hline
\hline
Reference & z & SN rate($10^{-4} Mpc^{-3} yr^{-1}$)  \\
\hline
Botticella \& al. (2008) & 0.30 & $0.22^{+0.10}_{-0.08}$ \\
\hline
Dahlen \& al. (2008) & 0.47 & $0.80^{+1.66}_{-0.27}$ \\
\hline
Dahlen \& al. (2008) & 0.83 & $1.30^{+0.73}_{-0.51}$ \\
\hline
Dahlen \& al. (2008) & 1.21 & $1.32^{+0.32}_{-0.38}$ \\
\hline
Dahlen \& al. (2008) & 1.61 & $0.42^{+0.39}_{-0.23}$ \\
\hline
Dilday \& al. (2008) & 0.09 & $0.29^{+0.09}_{-0.07}$ \\
\hline
Horesh \& al. (2008) & 0.2 & $0.189\pm 0.042$ \\
\hline
Kuznetsova \& al. (2008) & 0.4 & $0.53^{+0.39}_{-0.17}$ \\
\hline
Kuznetsova \& al. (2008) & 0.8 & $0.93\pm 0.25$  \\
\hline
Kuznetsova \& al. (2008) & 1.2 & $0.75^{+0.35}_{-0.30}$ \\
\hline
Kuznetsova \& al. (2008) & 1.55 & $0.12^{+0.58}_{-0.119}$ \\
\hline
Dilday \& al. (2010) & 0.12 & $0.269^{+0.034}_{-0.030}$ \\ 
\hline
Li \& al. (2011) & 0 & $0.301^{+0.062}_{-0.061}$ \\
\hline
Rodney \& Tonry (2010) & 0.15 & $0.32\pm 0.32$ \\
\hline
Rodney \& Tonry (2010) & 0.35 & $0.34\pm 0.19$ \\ 
\hline
Rodney \& Tonry (2010) & 0.45 & $0.31\pm 0.15$ \\ 
\hline
Rodney \& Tonry (2010) & 0.55 & $0.32\pm 0.14$ \\
\hline
Rodney \& Tonry (2010) & 0.65 & $0.49\pm 0.17$ \\
\hline
Rodney \& Tonry (2010) & 0.75 & $0.68\pm 0.21$ \\
\hline
Rodney \& Tonry (2010) & 0.85 & $0.78\pm 0.22$ \\ 
\hline
Rodney \& Tonry (2010) & 0.95 & $0.76\pm 0.25$ \\
\hline
Rodney \& Tonry (2010) & 1.05 & $0.79\pm 0.28$ \\
\hline
Graur \& al. (2011) & 0.74 & $0.79^{+0.33}_{-0.41}$ \\
\hline
Graur \& al. (2011) & 1.23 & $0.84^{+0.25}_{-0.28}$ \\
\hline
Graur \& al. (2011) & 1.69 & $1.02^{+0.54}_{-0.36}$ \\
\hline
\hline
\end{tabular}
\end{center}
\caption{A compilation of the observational cosmic Type Ia rates at different redshifts up to $z\sim1.75$.}
\label{dati1} 
\end{table*}

\end{document}